\pdfoutput=1
\documentclass{ws-ijgmmp}
\usepackage{amsmath,amsfonts}
\usepackage{array}

\newcommand{\ud}{\mathrm{d}}

\begin{document}

\markboth{M.~Szyd{\l}owski, O.~Hrycyna \& A.~Stachowski}{Scalar Field Cosmology -- Geometry of Dynamics}

\catchline{}{}{}{}{}

\title{SCALAR FIELD COSMOLOGY -- GEOMETRY OF DYNAMICS}

\author{MAREK SZYD{\L}OWSKI}
\address{Astronomical Observatory, Jagiellonian University, \\Orla 171,
30-244 Krak{\'o}w, Poland\\ Mark Kac Complex Systems Research Centre, Jagiellonian
University,\\ Reymonta 4, 30-059 Krak{\'o}w, Poland \\
\email{marek.szydlowski@uj.edu.pl}}
\author{OREST HRYCYNA}
\address{Theoretical Physics Division, National Centre for Nuclear
Research, \\ Ho{\.z}a 69, 00-681 Warszawa, Poland\\
\email{orest.hrycyna@fuw.edu.pl}}
\author{ALEKSANDER STACHOWSKI}
\address{Astronomical Observatory, Jagiellonian University, \\Orla 171,
30-244 Krak{\'o}w, Poland}

\maketitle

\begin{history}
\received{(Day Month Year)}
\revised{(Day Month Year)}
\end{history}

\begin{abstract}
We study the Scalar Field Cosmology (SFC) using the geometric language of the phase space. We define and study an ensemble of dynamical systems as a Banach space with a Sobolev metric. The metric in the ensemble is used to measure a distance between different models. We point out the advantages of visualisation of dynamics in the phase space. It is investigated the genericity of some class of models in the context of fine tuning of the form of the potential function in the ensemble of SFC.

We also study the symmetries of dynamical systems of SFC by searching for their exact solutions. In this context we stressed the importance of scaling solutions. It is demonstrated that scaling solutions in the phase space are represented by unstable separatrices of the saddle points. Only critical point itself located on two dimensional stable submanifold can be identified as scaling solution. We have also found a class of potentials of the scalar fields forced by the symmetry of differential equation describing the evolution of the universe. A class of potentials forced by scaling (homology) symmetries was given. We point out the role of the notion of a structural stability in the context of the problem of indetermination of the potential form of the Scalar Field Cosmology. We characterise also the class of potentials which reproduces the $\Lambda$CDM model, which is known to be structurally stable. We show that the structural stability issue can be effectively used is selection of the scalar field potential function. This enables us to characterise a structurally stable and therefore a generic class of SFC models. We have found a nonempty and dense subset of structurally stable models. We show that these models possess symmetry of homology.
\end{abstract}

\section{Introduction}

The matter content in the standard cosmology is usually described in terms of perfect fluid which satisfies a barotropic form of the equation of state $p= p(\rho) $, where $p$ and $\rho$ are functions of the cosmological time $t$ and represent the pressure and the energy density of the matter, respectively. The divergence of the energy momentum tensor, describing the source of gravity, is vanishing as a consequence of the Bianchi identities. Therefore, a general coordinate invariance of the general relativity and the theory of the general relativity applied in cosmology are self-consistent.

A very special form of the spacetime metric is postulated at the starting point in the standard cosmological model. The Fried\-mann-Robertson-Walker (FRW) metric is motivated by the observation which indicates that the Universe is homogeneous and isotropic at the very large scale (larger than 100 Mpc), i.e., the cosmological principle has the cosmological justification. As a consequence the evolution of the Universe is described in terms of a single function of the cosmic time, namely by the scale factor $a(t)$. This function satisfies the second order differential equation which admits the first integral in the form of the first order differential equation,
\begin{equation}
    3 H^2 = \rho_{\text{eff}} (a) = \rho_{m, 0} a^{-3} (t) + \Lambda - \frac{3 k}{a^2}\,,
\label{eq:fri}
\end{equation}
where $H= (\ln{a})^{.}$ (a dot means the differentiation with respect to $t$), $\Lambda$ is the cosmological constant, $k=0, \pm 1 $ is the curvature constant and the covariant condition of conservation of the energy-momentum tensor reduces to
\begin{equation}
    \dot{\rho} = -3 \frac{\dot{a}}{a} (\rho +p)\,.
\end{equation}

Equation \eqref{eq:fri} is called the Friedmann equation. There is a simple parallel with the motion of a particle of the unit mass in the potential $V=V(a)$
\begin{equation}
    V(a)=-\frac{\rho_{\text{eff}}(a)a^2}{6}\,.
\end{equation}
We have
\begin{equation}
    \frac{\dot{a}^2}{2} + V(a) = 0\,,
\end{equation}
on the zero energy level, where $\rho_{\text{eff}}$ plays the role of effective energy density parameterized through the scale factor $a(t)$. The cosmological constant as well as the curvature term in Eq.\eqref{eq:fri} can be interpreted formally as a special kind of fluid: $p_{\Lambda}=-\rho_{\Lambda}=-\Lambda$ as well as $p_k=-\frac{1}{3} \rho_k $, $\rho_k=-\frac{3k}{a^2}$.
Therefore the standard cosmological model can be simply represented in the terms of a dynamical system of a Newtonian type: $\ddot{a} = -\frac{\partial V}{\partial a}$, where the scale factor $a$ plays the role of a positional variable of a fictitious particle of the unit mass, which mimics the expansion of the Universe.

Note that, if we interpret the cosmological constant $\rho_\Lambda$ as an energy of a quantum vacuum, then this energy should remain constant during the whole cosmic evolution $(p_{\Lambda}=-\rho_{\Lambda})$ since $\dot \rho = - 3H(\rho + p)=0$.

An alternative method of describing the matter content of the Universe is to adopt the energy momentum tensor of a scalar field $\phi$ with the kinetic part and the potential $V=V(\phi)$. The energy momentum tensor is \cite{Faraoni:book}
\begin{equation}
    T_{\mu\nu} = \nabla_{\mu}\phi\nabla_{\nu}\phi -
\frac{1}{2}g_{\mu\nu}\nabla^{\alpha}\phi\nabla_{\alpha}\phi -
V(\phi)g_{\mu\nu}\,.
\end{equation}

If the FRW metric is postulated (cosmological principle is assumed) then the scalar fields  are homogeneous, i.e. $\phi = \phi (t)$. Formally the cosmology with the homogeneous scalar fields can be treated as a cosmology with perfect fluid; so the energy density and the pressure are
\begin{equation}
    \rho_{\phi} = \frac{{\dot{\phi}}^2}{2}+V(\phi),\ p_{\phi}= \frac{{\dot{\phi}}^2}{2}-V(\phi).
\end{equation}

Note that if the scalar field is slowly changing in time, then it is possible to reproduce the equation of state for the cosmological constant. However, in general the barotropic form of the equation of state $p=p(\rho)$ for the scalar field does not always exist in this special form.

There are many reasons to replace the ``paradigm'' of describing the matter content in terms of the barotropic equation of state by the description in terms of scalar field. Let us mention some of them :
\begin{arabiclist}
\item{The standard cosmological model requires the extension to the Planck epoch where quantum cosmology is dedicated to description of high energetic states of the Universe evolution. The barotropic equation of state seems to be unnatural within  this description and just scalar fields instead of a barotropic matter are preferred}
\item{The problem of vacuum energy in the standard cosmology (as well as probably in the contemporary particle physics) appeared. If we explain in cosmology an accelerating phase of an expansion of the Universe through the cosmological constant $\Lambda$ (which we interpret as vacuum energy) then theoretical expectations differ from the supernovae observation by 50-100 orders of magnitude. We are looking for some physical mechanism of the vacuum decay leading $\Lambda$ to the suitably small value which will be able to drive an acceleration of the current Universe}
\end{arabiclist}
The idea of quintessence is offered in this context as the solution of the cosmological constant problem, i.e., an answer to the question, why the value of the cosmological constant (vacuum energy) is so small in the current Universe. Instead of  $\Lambda(t)$, which is changing in time $t$ and thus becomes inconsistent with the energy conservation condition, we postulate the presence of a scalar field with the potential $V(\phi)$, minimally or non-minimally coupled to gravity. In order to satisfy the conservation of the energy momentum tensor, a massless scalar field is necessary. The condition of conservation of the energy-momentum tensor for the non-minimally scalar field in the FRW universe reduces to equation of motion (the Klein-Gordon equation)
\begin{equation}
    \ddot\phi+3H\dot\phi+U'(\phi, R)=0,
\label{eq:KG}
\end{equation}
where the term of non-minimal coupling to gravity was taken in the very simple form: $U=\xi R \frac{\phi^2}{2}$, $'\equiv\frac{\partial}{\partial \phi}$ denotes partial differentiation with respect the scalar field, $R$ denotes the Ricci scalar. If in eq. \eqref{eq:KG} the function $U$ is replaced by $V(\phi)$ then we obtain the form of the equation of motion for the scalar field minimally coupled to gravity.

Formally the effects of the non-minimal coupling constant $\xi$ can be treated also as an effect of a perfect fluid having an the energy density and a pressure in the following form 
\begin{align}
    \rho_\phi &= \epsilon\frac{\dot \phi^2}{2}+V(\phi)+3 \epsilon \xi H^2 \phi^2 + 3 \epsilon \xi H (\phi^2)^{.}, \\
    p_\phi &= \epsilon(1-4\xi)\frac{\dot\phi^2}{2}-V(\phi)+\epsilon\xi H (\phi^2)^{.}-2\epsilon\xi(1-6\xi)\dot H \phi^2 \nonumber \\
    & \;\;\;\; - 3\epsilon\xi(1-8\xi)H^2\phi^2,
\end{align}
where $\epsilon=\pm 1$ for canonical and phantom scalar fields, respectively.
One can see that in a very special case the equation of state for the scalar field with the potential only can be reduced to the form of $p=p(\rho)$. Of course, a Scalar Field Cosmology (SFC) model can mimic the Standard Cosmological Model (SCM) in this sense that SFC reproduces the same relation $H(a)$ like that obtained for SCM or reproduces the same value of a potential of a particle mimicking the evolution of the universe \cite{Astashenok:2012kb}. For SCM $H(a)$ relation has the form
\begin{equation}
    3 H^2(a)=\rho_{m,0}a^{-3} +\Lambda\,.
    \label{eq:10}
\end{equation}
In the case of SFC the above relation has the following form
\begin{align}
    3 H^2(a)&=\rho_{m,0}a^{-3}+\rho_\phi \nonumber \\
    & =\rho_{m,0}a^{-3}+\frac{1}{2}\left( \frac{d \phi}{d a}\right) ^2 H^2(a)a^2+V(\phi(a))\,,
    \label{eq:11}
\end{align}
where it is assumed that the scalar field $\phi$ depends only on the cosmic time $t$ through the scale factor, i.e. $\phi(t)=\phi(a(t))$.

Comparing $H^2(a)$ relations \eqref{eq:10} and \eqref{eq:11} both, which identify uniquely the cosmic evolution (which is probing also by cosmography methods), we obtain:
\begin{equation}
    \rho_{m,0}a^{-3}+\Lambda=\frac{\rho_{m,0}a^{-3}+V(\phi(a))}{1-\frac{1}{6}(\frac{d\phi}{d a})^2 a^2},
\end{equation}
where $\phi$ and $V(\phi)$ satisfy the Klein-Gordon equation: $\ddot\phi+3H\dot\phi+V'(\phi)=0$.

In the general if we assume the matter content in terms of the scalar field this means that we are going beyond the standard cosmological model. We will assume in our future investigations the presence of both kinds of the matter content described by the barotropic form of the equation of state and by the scalar field. In such a case one can discover new interesting effects which are not present in the cosmological evolution if a scalar field as well as the matter are treated separately \cite{Hrycyna:2008gk,Hrycyna:2012tn}.

In this scalar field ``paradigm'' some problems related with indetermination of the potential of the scalar field, appear in a description of the matter content. This indetermination is of the same type as in SCM where the equation of state is taken from local physics and is postulated before performing the integration of the Einstein equation (see the Friedmann equation where the energy density appears as a function of the scale factor $a$). The problem is which potential for the cosmological scalar field should be taken into account? Will the evolutional scenarios be sensitive on the choice of a form the potential? This question is very important for the consistency of SFC.

Some forms of potential (for example quadratic, exponential, etc.) are very popular in SFC but there are not theoretical or observational reasons indicating which of them should determine uniquely our choice? With this before addressed problem another one is connected: how to treat or to interpret the Klein-Gordon equation? Is it an equation of motion for scalar field or it is just dedicated for a determination of the potential function \cite{Chervon:1997yz}?

There is progress in the searching for analytical solutions in the SFC. Some interesting results were obtained by Kamenschik group \cite{Andrianov:2011fg} and others \cite{Chervon:1997yz}.

The main aim of this paper is to study the evolution of the Universe using dynamical system methods as well as methods of a symmetry
for an investigation geometric and algebraic structures both in the ensemble of the scalar field cosmological model.

\section{Scalar Field Cosmology as a dynamical system}

\subsection{Cosmology as an effective theory of the Universe}

If we look at the modern cosmology and particle physics we can see similarities in their methodologies. Moreover many crucial problems of contemporary cosmology lie on the intersections of cosmology and particle physics. It is so because similarly to the  particle physics the modern cosmology has a methodological status of an effective theory. In both theories there is a crucial notion of `standard model' which is sometimes called the concordance model in cosmology.

The characteristic features of the standard cosmological model are the following:
\begin{arabiclist}
\item{The theory possesses some cuts-off or energetic cuts which reveal limits of the classical singularity. Describing the structure and evolution of the Universe we apply the classical theory of the general relativity and Planck epoch does not belong to the domain of our interest}
\item{To investigate global properties of the Universe we use some parameters (Density parameters and some parameters related with the spectrum of CMB radiation play the role of such parameters in the cosmology). The nature of some parameters is unknown and we only believe that a new more fundamental theory will give us the explanation of their nature: A connection of the density parameter with the cosmological constant is an example of such a case. Other parameters are fitted from the astronomical observations. In general the parameters used in the model may be useful `fictions' in its construction which plays crucial role in the investigation and construction of the model of the Universe. Some authors claim that density parameters are dressed (see, eg. \cite{Buchert:2002ij})}
\item{The main purpose of our researches is to construct such a consistent model of the Universe which is reliable, effective and good working: It should give possibility to define new observables. Within such a scheme using the effective theory one can extrapolate the physics to the new domains}
\end{arabiclist}
The example of such effective theories of the Universe are SFC models. We do not know what are they scalar fields but this notion is a useful instrument for a description of the matter and allows one to predict some new phenomena.

\subsection{Evolution of the Universe in the phase space.}

The Einstein field equations of general relativity constitute a very complicated system of partial and nonlinear equations and without some simplified assumptions the problem of solving them is very complex. From the cosmological point view there is an important subclass of models in a broad class of solutions of Einstein  equations which can be reduced to the dynamical system,
\begin{equation}
    \frac{\ud x}{\ud t}=f(x),
\end{equation}
where $f$ is a function of class $C^1$ on M and $x\in R^n$, $x$ is a vector of state variables and $M$ is a open subset of $R^n$.

If a symmetry of the space is postulated (eg. as that following from the cosmological principle (without isotropy)) then the Einstein equations reduces to the form of ordinary differential equation, i.e., to a dynamical system of Bianchi models.

The possibility of investigations of dynamical systems
for all admissible initial conditions is the
main advantage of representing the cosmic evolution in terms of such systems. It is especially important in cosmology because initial conditions for the Universe are unknown. In some sense the cosmology is a physical theory of the Universe as well as a theory of initial conditions which lead to the current Universe.

Moreover methods of the dynamical systems allow us to analyse the stability of evolutional paths in the phase space.

The right hand sides of the dynamical system represent vector fields in the tangent space of phase space. In general a vector field depends on the set of model parameters. If the values of parameters are fixed then we can ask about the global phase portrait which is geometrical visualisation of whole global dynamics of a system. One can also study how this global dynamics changes when the model parameters vary. This question can be explained with the help of the bifurcation theory. For example, it is interesting to find how the the global phase portraits change when the non-minimal coupling constant is included into model. The equivalence of phase portraits is established by modulo homeomorphism transforming the phase trajectories of both system and preserving a direction of time along the trajectories. If the qualitative behaviour of the system $\dot x=f(x)$ remains equivalent in the topological sense of a homeomorphism preserving the orientation of a phase curve, then one can say that systems are equivalent.

To investigate the stability of the solution, we linearize the system around the fixed point (a solution of the system for which all right hand sides vanish). If this point, called also a critical point, is non-degenerate (all eigenvalues of a linearization matrix are nonzero) then following the Hartman-Grobman theorem \cite{Perko:2001} the linear part of the system  is a well approximation of the system in the neighbourhood of this point. Therefore, for non-degenerate critical points, the construction of phase portraits reduces to the study of fixed points and its stability in the terms of eigenvalues of the linearization matrix.

\subsection{Scalar field cosmology with minimally and a non-minimally coupled scalar field in the phase space}

It is assumed in the scalar field cosmology with the non-minimal coupling to gravity of type $\frac{1}{2}\xi R\phi^2$ that the action of the gravity and matter has the following form
\begin{equation}
	S=\frac{1}{2}\int \ud^4 x \sqrt{-g} \left( R-\epsilon(g^{\mu\nu}\partial_\mu\phi\partial_\nu\phi+\xi R \phi^2)-2V(\phi)\right)+S_m.
\end{equation}
The equations of motion for the scalar field are obtained after a variation of the action under the scalar field. In such a way one obtains
\begin{equation}
    \ddot\phi+3H\dot\phi+\xi R\phi+\epsilon V'(\phi)=0,
\end{equation}
where $\xi$ is the non-minimal coupling constant, $R$ is the Ricci scalar and $\epsilon=\pm 1$ for the canonical and phantom scalar fields, a prime denotes differentiation with respect to $\phi$, $'\equiv\frac{\ud}{\ud\phi}$.
Performing a variation under the FRW metric we obtain the FRW equation. Its first integral has the following form,
\begin{equation}
    E=\epsilon\frac{1}{2}\dot\phi^2+\epsilon 3 \xi H^2 \phi^2+\epsilon 3 \xi H(\phi^2)^. + V(\phi)+\rho_m-3H^2,
\end{equation}
or
\begin{equation}
    3H^2=\rho_\phi+\rho_m,
\end{equation}
for a flat universe.
The acceleration equation looks as follows
\begin{equation}
    \dot H=-\frac{1}{2}\big((\rho_\phi+p_\phi)+\rho_m (1+w_m)\big),
\end{equation}
where
\begin{equation}
\begin{split}
    & \rho_{\phi} = \epsilon\frac{1}{2}\dot \phi^2+V(\phi)+\epsilon 3 \xi H^2\phi^2+\epsilon 3 \xi H(\phi^2)^. \,,\\
    & p_{\phi} = \epsilon\frac{1}{2}(1-4\xi)\dot\phi^2-V(\phi)+\epsilon\xi H(\phi^2)^.-\epsilon 2 \xi (1-6\xi)\dot H \phi^2 - \\ &\qquad
-\epsilon 3 \xi (1-8\xi)H^2 \phi^2+2 \xi \phi V'(\phi),
\end{split}
\end{equation}
and $w_m$ is a coefficient in the equation of state for the matter $p_{m}=w_{m} \rho_{m}$.

It is convenient to introduce new state variables
\begin{equation}
	x\equiv\frac{\dot{\phi}}{\sqrt{6} H},\ y\equiv\frac{\sqrt{V(\phi)}}{\sqrt{3} H},\ z\equiv\frac{1}{\sqrt 6}\phi
\end{equation}
which represent a state of the system in the phase space.

The acceleration equation can be rewritten in a new form in which a fluid with the effective energy density $\rho_{\text{eff}}$ and an effective pressure $p_{\text{eff}}$ appear
\begin{equation}
    \dot{H}=-\frac{1}{2}(\rho_{\text{eff}}+p_{\text{eff}})=-\frac{3}{2}H^2(1+w_{\text{eff}}),
\end{equation}
where
\begin{equation}
\begin{split}
    & w_{\text{eff}}=\frac{p_{\text{eff}}}{\rho_{\text{eff}}}= \\ & \frac{1}{1-\epsilon 6 \xi (1-6 \xi) z^2}\Big(-1+\epsilon(1-6\xi)(1-w_m)x^2+ \\ & \quad + \epsilon 2 \xi (1-3 w_m)(x+z)^2+(1+w_m)(1-y^2)-\epsilon 2 \xi (1-6\xi)z^2-2\xi\lambda y^2 z\Big),
\end{split}
\end{equation}
where $\lambda=-\sqrt 6\frac{1}{V(\phi)}\frac{dV(\phi)}{d\phi}$ is a quantity characterising geometry of the potential. The dynamics of the model under consideration can be represented as the following 4-dimensional dynamical system
\begin{equation}
\begin{split}
    & x'=-(x-\epsilon\frac{1}{2}\lambda y^2)\big(1-\epsilon 6 \xi(1-6\xi)z^2\big)+ \\ & \qquad +\frac{3}{2}(x+6\xi z)\bigg(-\frac{4}{3}-2\xi\lambda y^2 z+\epsilon(1-6\xi)(1-w_m)x^2+ \\ & \hspace{3cm} + \epsilon 2 \xi (1-3 w_m)(x+z)^2+(1+w_m)(1-y^2)\bigg)\,,\\
    & y'=y(2-\frac{1}{2}\lambda x)\big(1-\epsilon 6 \xi (1-6\xi)z^2\big)+ \\ & \qquad + \frac{3}{2}y\Big(-\frac{4}{3}-2 \xi y^2 z+\epsilon(1-6\xi)(1-w_m)x^2+ \\ & \hspace{2.5cm} + \epsilon 2 \xi (1-3w_m)(x+z)^2+(1+w_m)(1-y^2)\Big)\,,\\
   & z'=x\big(1-\epsilon 6 \xi (1-6 \xi)z^2\big)\,,\\
   & \lambda'=-\lambda^2(\Gamma-1)x\big(1-\epsilon 6 \xi (1-6\xi)z^2\big),
\end{split}
\end{equation}
where $\Gamma=\frac{\frac{d^2 V(\phi)}{d\phi^2}V(\phi)}{(\frac{dV(\phi)}{d\phi})^2}$ and original time variable is replaced by a new time parameter $\tau$,
\begin{equation}
    \frac{\ud}{\ud\tau}=\big(1-\epsilon 6 \xi (1-6\xi)z^2\big)\frac{\ud}{\ud \ln a}.
\end{equation}
If we assume, for simplicity, that $\Gamma=\Gamma(\lambda)$ then the last two equations can be directly integrated. Then
\begin{equation}
\frac{\ud z(\lambda)}{\ud\lambda}=z'(\lambda)=-\frac{1}{\lambda^2(\Gamma(\lambda)-1)}.
\end{equation}
Therefore
\begin{equation}
    z(\lambda)=-\int\frac{\ud\lambda}{\lambda^2(\Gamma(\lambda)-1)}.
\end{equation}
Let us assume the following parameterisation of the function $\Gamma(\lambda)$ which covers the large class of potentials (see Table \ref{tab:1})
\begin{equation}
    \Gamma(\lambda)=1-\frac{1}{\lambda^2}(\alpha+\beta\lambda+\gamma\lambda^2).
\end{equation}

\begin{table}
\tbl{Examples of potential functions for various configurations of values of parameters as well as of the assumed form of the $\Gamma(\lambda)$.\label{tab:1}}
{\begin{tabular}{@{}ccc@{}} \toprule
parameters & $z(\lambda)$ & $V(\phi)$ \\ \colrule
 $\alpha\neq 0$, $\beta=0$, $\gamma=0$ & $\frac{\lambda}{\alpha}+\text{const}$ & $V_0\exp(-\frac{\alpha}{2}\phi^2+\text{const} \,\phi)$\\
 $\alpha= 0$, $ \beta\neq 0$, $\gamma=0$ & $\frac{\ln\lambda}{\beta}+\text{const}$ & $V_0\exp(\frac{\text{const}}{\beta}\exp(\beta \phi))$\\
 $\alpha= 0$, $\beta= 0$, $ \gamma\neq 0$ & $-\frac{1}{\gamma\lambda}+\text{const}$ & $V_0(\gamma\phi-\text{const})^{\frac{1}{\gamma}}$\\
 $\alpha\neq 0$, $\beta\neq 0$, $\gamma= 0$ & $\frac{\ln(\alpha+\beta\lambda)}{\beta}+\text{const}$ & $V_0\exp(\frac{1}{\beta}(\alpha\phi+\text{const}\,\exp(\beta\phi)))$\\
 $\alpha\neq 0$, $\beta= 0$, $\gamma\neq 0$ & $\frac{\arctan(\sqrt{\frac{\gamma}{\alpha}}\lambda)}{\sqrt{\alpha\gamma}}+\text{const}$ & $V_0(\cos(\sqrt{\alpha\gamma}(\phi-\text{const})))^{\frac{1}{\gamma}}$\\
 $\alpha= 0$, $\beta\neq 0$, $\gamma\neq 0$ & $ -\frac{\ln \lambda-\ln(\beta+\gamma\lambda)}{\beta}+\text{const}$ & $ V_0(\exp(\text{const}\,\beta)+\gamma\exp(\beta\phi))^{\frac{1}{\gamma}}$\\
 $\alpha\neq 0$, $ \beta\neq 0$, $\gamma\neq 0$ & $\frac{2\arctan\left(\frac{\beta+2\gamma\lambda}{{\sqrt{-\beta^2+4\alpha\gamma}}}\right)}{\sqrt{-\beta^2+4\alpha\gamma}}+\text{const}$ & $ V_0\exp(\frac{\beta}{2\gamma}\phi)(\cos(\frac{1}{2}\sqrt{-\beta^2+4\alpha\gamma}(\phi-\text{const})))^{\frac{1}{\gamma}}$\\
 \botrule
  \end{tabular}}
  \end{table}
There are different choices of $\Gamma(\lambda)$ in Table \ref{tab:1} and corresponding to them a form of their potential. Making choice, e.g., $\Gamma(\lambda)$ in the form
\begin{equation}
\Gamma(\lambda)=\frac{3}{4}-\frac{\sigma^2\lambda^2}{4(2+\sqrt{4\pm\sigma^2\lambda^2})^2}
\end{equation}
we get $z(\lambda)$ in the form
\begin{equation}
    z(\lambda)=-\frac{2+\sqrt{4\pm\sigma^2\lambda^2}}{\lambda}+ \text{const}
\end{equation}
which reproduces the Higgs potential
\begin{equation}
	V(\phi)=V_0 ((\phi- \text{const})^2-\sigma^2)^2.
\end{equation}
For $z=z(\lambda)$ the dynamical system takes the following form
\begin{equation}
\label{eq:dynsys}
\begin{split}
   & x'=-(x-\epsilon\frac{1}{2}\lambda y^2)\big(1-\epsilon 6 \xi (1-6\xi)z(\lambda)^2\big)+\\ & \qquad +\frac{3}{2}(x+6\xi z(\lambda))\bigg(-\frac{4}{3}-2\xi\lambda y^2 z(\lambda)+\epsilon(1-6\xi)(1-w_m)x^2+\\ & \hspace{3.5cm}+\epsilon 2 \xi (1-3 w_m)(x+z(\lambda))^2+(1+w_m)(1-y^2)\bigg)\,,\\
   & y'=y(2-\frac{1}{2}\lambda x)\big(1-\epsilon 6 \xi (1-6\xi)z(\lambda^2)\big)+ \\ & \qquad 
   +\frac{3}{2}y\bigg(-\frac{4}{3}-2\xi\lambda y^2 z(\lambda)+\epsilon(1-6\xi)(1-w_m)x^2+ \\ & \hspace{2.5cm} +\epsilon 2 \xi(1-3w_m)(x+z(\lambda))^2+(1+w_m)(1-y^2)\bigg)\,,\\
   & \lambda'=-\lambda^2 (\Gamma(\lambda)-1)x\big(1-\epsilon 6 \xi(1-6\xi)z(\lambda)^2\big)\,.
\end{split}
\end{equation}
Critical points of this 3-dimensional system are completed in Table \ref{tab:2}.

\begin{table}
\tbl{Localization of critical points $(x^{*}, y^{*}, z^{*})$ of the system \eqref{eq:dynsys} and corresponding values of $w_{\text{eff}}$.\label{tab:2}}
{\begin{tabular}{cccc} \toprule
   $x^{*}$  & $y^{*}$ &  $\lambda^{*}$ & $w_{\text{eff}}$ \\  \colrule
  $x^{*}_{1}=-6\xi z(\lambda^{*}_{1})$ & $ y^{*}_{1}=0$ & $\lambda^{*}_{1} \colon z(\lambda)^{2}=\frac{1}{\epsilon 6\xi(1-6\xi)}$ & $\pm\infty$ \\
 $x^{*}_{2a} = -6\xi z(\lambda^{*}_{2a})$ &
  $(y^{*}_{2a})^{2}=\frac{4\xi}{2\xi\lambda^{*}_{2a}z(\lambda^{*}_{2a})+(1+w_{m})}$ & 
  $\lambda^{*}_{2a} \colon z(\lambda)^{2} = \frac{1}{\epsilon 6\xi(1-6\xi)}$ & $w_{m}-4\xi$ \\
 $x^{*}_{2b} = 0$ & $(y^{*}_{2b})^{2} = \frac{2\xi(1-3w_{m})}{(1-6\xi)\big(2\xi\lambda^{*}_{2b}z(\lambda^{*}_{2b})+(1+w_{m})\big)}$
& $\lambda^{*}_{2b} \colon z(\lambda)^{2} = \frac{1}{\epsilon 6\xi(1-6\xi)}$ &
$\frac{w_{m}-2\xi}{1-6\xi}$ \\
 $x^{*}_{3a}:g(x)=0${\scriptsize $^{1}$}
  & $y^{*}_{3a}=0$ & $\lambda^{*}_{3a} \colon z(\lambda)^{2} = \frac{1}{\epsilon 6\xi(1-6\xi)}$ & $\frac{1}{3}$ \\
 $x^{*}_{3b}=0$ & $y^{*}_{3b}=0$ & $\lambda^{*}_{3b} \colon z(\lambda)^{2} = \frac{1}{\epsilon 6\xi}$ & $\frac{1}{3}$ \\
$x^{*}_{4} = 0$ & $y^{*}_{4} = 0$ & $\lambda^{*}_{4} \colon z(\lambda)=0$
& $w_{m}$ \\
 $x^{*}_{5} = 0$ & $(y^{*}_{5})^{2} = 1 - \epsilon 6\xi z(\lambda^{*}_{5})^{2}$
& $ \lambda^{*}_{5} \colon \lambda z(\lambda)^{2} + 4 z(\lambda)-\frac{\lambda}{\epsilon 6\xi} = 0$ & $-1$ \\
  \botrule
  \end{tabular}}
{\scriptsize $^{1}g(x)=\epsilon (1-4\xi-w_{m})x^{2}+\epsilon 4\xi(1-3w_{m})z(\lambda^{*}_{3a})x+\frac{2\xi}{1-6\xi}(1-3w_{m})$}
\end{table}

Let us discuss now a nature of these critical points in details. One can distinguish the following critical points:
\begin{romanlist}[(iii)]
\item[(1)]{Finite scale factor singularity eigenvalues are $l_1=6\xi$, $l_2=6\xi$, $l_3=12 \xi$. The critical point is an unstable node for $\xi>0$, and it is a stable node for $\xi<0$}
\item[(2a)]{Fast-roll inflation eigenvalues are $l_1=0$, $l_2=12\xi$, $l_3=-12\xi$ and there is the non-hyperbolic critical point}
\item[(2b)]{Slow-roll inflation eigenvalues are $l_1=l_2=l_3=0$ and there is the degenerated critical point}
\item[(3a)]{The radiation domination epoch, which is generated by the non-minimal coupling for the phantom scalar field with $\xi>0$. Eigenvalues are $l_1=0$, $l_2>0$, $l_3<0$ and there is the non-hyperbolic critical point}
\item[(3b)]{The radiation domination epoch, which is generated by the non-minimal coupling for the canonical scalar field with $\xi>0$. Eigenvalues are $l_1=6\xi(1-3w_m)$, $l_2=12\xi$, $l_3=-6\xi$}
\item[(4)]{The matter domination epoch. Eigenvalues are the following
\begin{equation}
    l_{1,2}=-\frac{3}{4}\left((1-w_m)\pm\sqrt{(1-w_m)^2-\frac{16}{3}\xi(1-3w_m)}\right),\hspace{1mm}
    l_3=\frac{3}{2}(1+w_m)\,,
\end{equation}
and there is the non-degenerated point for $w_m\neq-1$ and $w_m\neq\frac{1}{3}$}
\item[(5)]{The present accelerated expansion epoch described by the de Sitter stage.
In the most general case, without assuming any specific form of the potential function we are unable to find coordinates of this point in exact form. In spite of this we are able to formulate general conditions for stability of this critical point. This requires that the real parts of the eigenvalues of the linearization matrix calculated at this point must be negative. From the Routh-Hurwitz criterion it follows that to assure stability of this critical point the following conditions should be fulfilled 
\begin{equation}
Re{[l_{1,2,3}]}<0 \iff 3\xi\frac{h'(\lambda^{*}_{5})}{z'(\lambda^{*}_{5})}
(y^{*}_{5})^{2} >0,
\end{equation}
where
$h(\lambda)= \lambda z(\lambda)^{2} +4 z(\lambda) -\frac{\lambda}{\epsilon 6\xi}$}\,.
\end{romanlist}

\begin{figure}
\begin{center}
\includegraphics[scale=0.8]{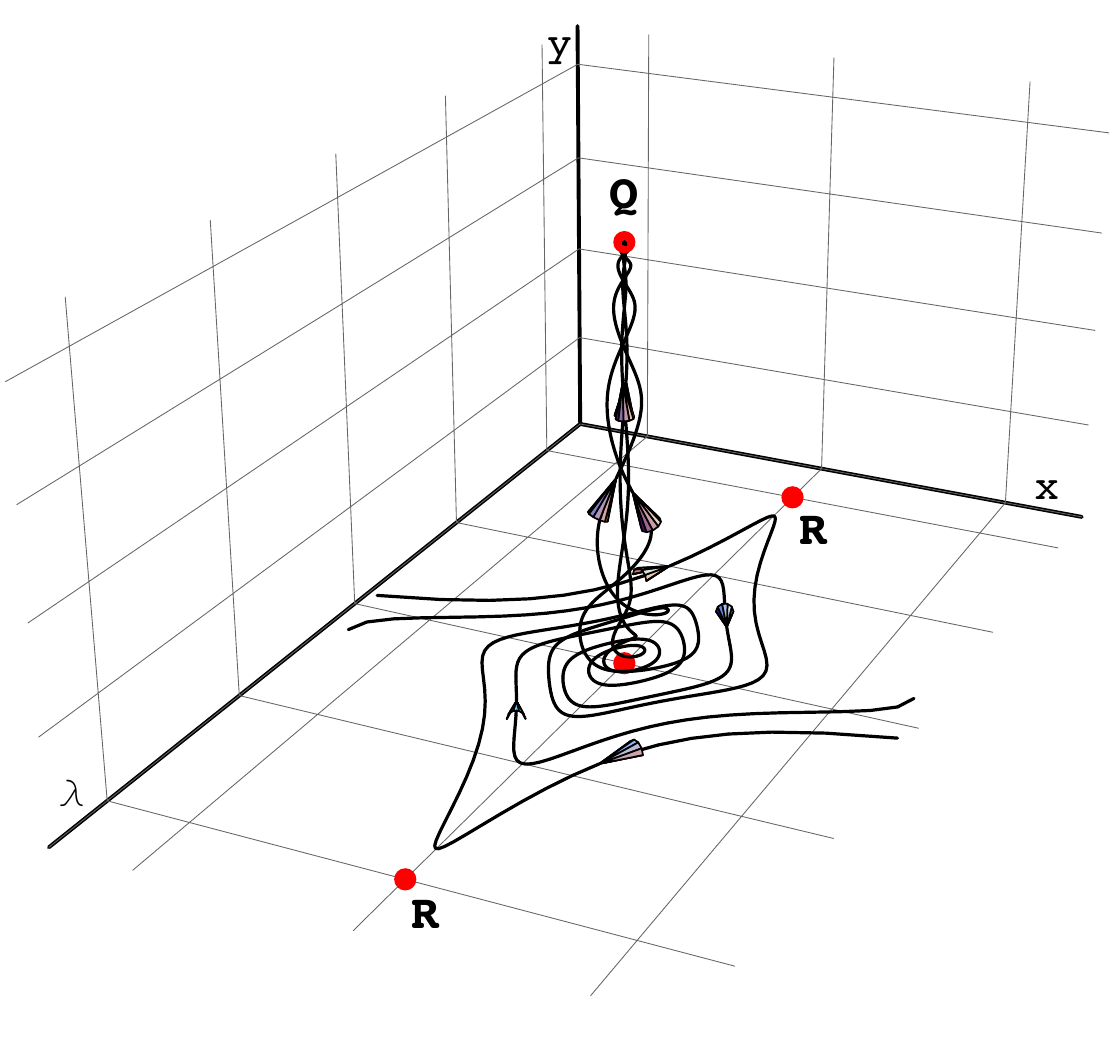}
\end{center}
\caption{The 3-dimensional phase portrait of the dynamical system \eqref{eq:dynsys} for $\Gamma(\lambda)=\frac{\lambda}{\alpha}$. Trajectories represent a twister type solution which interpolates between the radiation dominated universe (a saddle type critical point), the matter dominated universe (an unstable focus critical point) and the accelerating universe (a stable focus critical point).}
\label{fig:1}
\end{figure}

In the interesting example of the generic solution in the scalar field cosmology with the non-minimal coupling the dynamical system \eqref{eq:dynsys} can be represented in the phase space by trajectories which we call \textit{"a twister behavior"} \cite{Hrycyna:2009zj} (see the phase portrait in Fig.~\ref{fig:1}). It is interesting that this trajectory is going from the singularity of the finite scale factor toward the de Sitter attractor through the phases of a radiation and the matter dominating epoch. This type of behavior is generic in this sense that it does not depend on small changes of initial conditions.

\section{Ensemble of Scalar Field Cosmological Models (SFC)}

Let us define the notion of all cosmological dynamical system representing the evolution of the Universe filled by the matter and scalar field.

\begin{definition}[Ensemble of SFC models]
By ensemble $\mathcal{E}$ of homogeneous and isotropic cosmological models (i.e. FRW models) we understand space of all dynamical systems $\dot x=f(x)$, where $x\in R^n $ is state variable and dot denotes differentiation with respect time $t$ (or its monotonic function: $t \rightarrow \tau \colon dt=\Psi(x)d\tau$; $\Psi$ is $C^\infty$ class). A vector field $f$ is $C^1$ vector field defined on the open subset $E$ of $R^n $; $x$ is a positional variable which satisfies the Einstein field equation with the source of gravity in the form of the matter and scalar field, minimally and non-minimally coupled to gravity.
\end{definition}

The equivalence of ensembles establish the following definition:

\begin{definition}[Equivalence of ensembles]
We say that two ensembles $\mathcal{E}$ and $\mathcal{E}'$ are equivalent if there exists the homeomorphism $h$ transforming the vector field, say $f(x)$ into $g(x)$ (of the same dimension) with the orientation preservation of the phase curve. Of course, if there is a diffeomorphism transforming integral curves of $f(x)$ into those of $g(x)$ then corresponding ensembles we treated as equivalent.
\end{definition}

Using the last definition one can introduce equivalence classes and thus the space of all dynamical system in $R^n$ can be divided into disjoint classes of abstractions (equivalence classes).

If we think about elements of an ensemble in this paper we will use a hidden assumption that we are dealing only with the representants of this equivalence relation. The vector field becomes a function of state variables in cosmological applications and it can also depend on the model parameters.

It is interesting to know how some classes of dynamical systems of a cosmological origin are distributed in the ensemble. Are they generic in the ensemble $\mathcal{E}$ in the sense that they form an open and a dense subset? If they are generic then there is a strong theoretical argument for choosing these models (and a potential of the scalar field) and thus, for example, we overcome some indetermination in SFC related with choosing a form of the potential.

The notion of a structural stability for the simple 2-dimensional dynamical system is useful in searching how cosmological models of some classes are ``distributed'' in the phase space. In particular using this notion one can answer the question: are they generic or not? Moreover if they are non-generic then one can propose different scenarios leading to generic one through introducing  a ``$\epsilon$-perturbation''. In such a case theory of bifurcation may be useful.

The idea of the structural stability was introduced by Andronov and Pontriagin in 1937 \cite{Andronov:1937}. This notion is a property of the system itself and characterises how stable is a global dynamics under a small perturbation.

If $f\in C^1 (M)$, where $M$ is an open subset of $R^n $, then $C^1$ norm of vector field can be introduced in standard way,
\begin{equation}
    ||f||_1=\sup_{x\in E} |f(x)| + \sup_{x\in E}||D f(x)||,
\end{equation}
where $|| \ldots ||$ denotes the Euclidean norm in $R^n$ and a usual norm of the Jacobi matrix $Df(x)$, respectively.

It is important for our consideration that a set of vector fields bounded in the $C^1$ norm forms the Banach space \cite{Perko:2001}. Using this norm we can measure the distance between any two dynamical system. Taking into account the aim of the paper it will be useful to introduce another norm in the ensemble $\xi$.

\begin{definition}[Norm in ensemble $\mathcal{E}$]
Let us consider some compact subset $K$ of $M$, then $C^1$ norm of vector field $f$ on $K$ can be defined as
\begin{equation}
    ||f||_1=\max|f(x)|+\max|Df(x)|<\infty.
\end{equation}
\end{definition}

\begin{definition}[$\epsilon$-perturbation of dynamical system]
Let $E=R^n $, then $\epsilon$-perturbation of vector field $f$ is the vector field $g\in C^1(M)$ satisfying $||f-g||_1<\epsilon$.
\end{definition}

\begin{definition}[Structural stability of the vector field]
A vector field $f\in C^1(M)$ is called to be structurally stable if there exists an $\epsilon>0$ such that for all $g\in C^1(M)$ with $||f-g||_1<\epsilon$, $f$ and $g$ are topologically equivalent on an open subset $R^n$.
\end{definition}

One can see that the global dynamics represented by the phase portrait, or equivalently, the vector field is important in Andronov and Pontriagin's definition of the structural stability.

An opinion was originally widespread in the scientific folklore that all realistic models of a physical reality should be structurally stable (like for realistic damped oscillators). Therefore scientists modelling the world in different scales believe that structural stability is a typical attribute of dynamical system modelling a physically realistic problem in science.

However since Smale's construction \cite{Smale:book} we know that it cannot be true for higher dimensional dynamical systems ($n\geq 3$). A three-body model system is a typical example of such a model which is structurally unstable but represents a model of physical processes.

The case of an ensemble of two-dimensional dynamical systems is very special because the Peixoto theorem gives us a characterisation of the structurally stable vector fields on a compact, two-dimensional manifold which are generic. If a vector field $f\in C^1(M)$ is not structurally stable it belongs to the bifurcation set $C^1(M)$. For such systems the global phase portrait dramatically changes when the vector field passes through a point of the bifurcation set \cite{Kurek:2007tb}.

\section{Structural stability property of SFC models}

\subsection{Structural stability of SFC models}

Using the theory of dynamical systems we focus attention rather on the study of the properties of global portraits than on investigations of individual solutions (phase space trajectories). The notion of an ensemble $\mathcal{E}$ appears in this context as a natural way for studying how generic are cosmological models used for solving different cosmological problems.

Our models are only an effective description of the reality and because of this we believe that taking into account the ensemble of them they are generic. We understand that they are only an approximation and therefore, we understand that they can not be attributed to only individual (non-generic) subset of the ensemble.

Our theoretical presumption of methodological nature is that all models are rather generic than a fine tuning and therefore non-typical in the ensemble.

In this section we will study dynamical systems on the plane. In this case there is simple way to introducing the metric $d$ in the ensemble $\mathcal{E}$ of all dynamical systems on the plane
\begin{equation}
\label{eq:met_d}
\begin{split}
    d(\mathbf{X}, \mathbf{Y})=\max\Big\{ & \sup_{(x,y)\in U_{i}}\Big|f_{\mathbf{X}}(x,y)-f_{\mathbf{Y}}(x,y)\Big|\,,\\
    & \sup_{(x,y)\in U_{i}} \Big|g_{\mathbf{X}}(x,y)-g_{\mathbf{Y}}(x,y)\Big|\,,\\
     & \sup_{(x,y)\in U_{i}}\Big|\frac{\partial f_{\mathbf{X}}(x,y)}{\partial x}-\frac{\partial f_{\mathbf{Y}}(x,y)}{\partial x}\Big|\,,\\
     & \sup_{(x,y)\in U_{i}}\Big|\frac{\partial g_{\mathbf{X}}(x,y)}{\partial x}-\frac{\partial g_{\mathbf{Y}}(x,y)}{\partial x}\Big|\,,\\ 
     & \sup_{(x,y)\in U_{i}}\Big|\frac{\partial f_{\mathbf{X}}(x,y)}{\partial y}-\frac{\partial f_{\mathbf{Y}}(x,y)}{\partial y}\Big|\,,\\
     & \sup_{(x,y)\in U_{i}}\Big|\frac{\partial g_{\mathbf{X}}(x,y)}{\partial y}-\frac{\partial g_{\mathbf{Y}}(x,y)}{\partial y}\Big|\,, i=1,\dots,k\,.\Big\},
\end{split}
\end{equation}
where $U_1,...,U_k$ are open sets covering $M\in \mathbf{R}^2$ \cite{Smale:1969,Smale:book} and dynamical systems have the form $\mathbf{X} \colon \frac{\ud x}{\ud t}=f_{\mathbf{X}}(x,y), \frac{\ud y}{\ud t}=g_{\mathbf{X}}(x,y)$ and $\mathbf{Y} \colon \frac{\ud x}{\ud t}=f_{\mathbf{Y}}(x,y), \frac{\ud y}{\ud t}=g_{\mathbf{Y}}(x,y)$.

\begin{definition}[Closeness of a planar dynamical system]
Two dynamical systems
\begin{equation}
\begin{split}
    \mathbf{X} & \colon \frac{\ud x}{\ud t}=f_{\mathbf{X}}(x,y), \frac{\ud y}{\ud t}=g_{\mathbf{X}}(x,y)\,,\\
    \mathbf{Y} & \colon \frac{\ud x}{\ud t}=f_{\mathbf{Y}}(x,y), \frac{\ud y}{\ud t}=g_{\mathbf{Y}}(x,y)\,
\end{split}
\end{equation}
are close when $f(x,y)$ and $g(x,y)$ are uniformly close along with their first derivatives.
\end{definition}

\begin{definition}[Structural stability]
A system of differential equations is structurally stable if every sufficiently close system (in the sense of metric \eqref{eq:met_d}) has the same phase portrait.
\end{definition}

The class of planar dynamical systems is very particular because our model is very simple. Nevertheless such a simple models appear in physics and cosmology where they play very important roles (for example they can play the role of toys models). The ``true'' system, which we are looking for, does not coincide with the model under consideration. We do not know which is the true system. Therefore, the results of our investigations of the model system will make sense if we can transfer them to the real system. Because we do not know the true system exactly, our dynamical analysis will be reliable only if the model system is structurally stable.

The Peixoto theorem characterises dynamical system on the plane and unfortunately it cannot be generalised to more dimensional case. Following this theorem, structurally stable systems on the compact space form an open and dense subset in the space of all dynamical systems i.e. they are generic.

Right-hand sides of dynamical system are given in the polynomial form in many cosmological applications. In such a case it is possible to construct the Poincar{\'e} sphere \cite{Perko:2001}. Within this approach one can project trajectories from the centre of a unit sphere
\begin{equation}
    S^2=\{(X,Y,Z) \colon X^2+Y^2+Z^2=1\},
\end{equation}
onto the $(x,y)$ plane tangent to $S^2$ at either the north or south pole \cite{Perko:2001}. Due to this construction which was introduced by Poincar{\'e} the critical points of the dynamical system at infinity are spread out along the equator. If we project the upper hemisphere $S^2$ onto $(x,y)$ plane then
\begin{equation}
    x=\frac{X}{Z}, \quad y=\frac{Y}{Z}
\end{equation}
or
\begin{equation}
    X=\frac{x}{\sqrt{1+x^2+y^2}}, \quad Y=\frac{y}{\sqrt{1+x^2+y^2}}, \quad Z=\frac{z}{\sqrt{1+x^2+y^2}}.
\end{equation}
This construction gives us a simple method of studying the behaviour of the system at infinity. The phase space becomes compact by adding the circle at infinity, i.e., it is the projective plane.

The Andronov-Pontriagin theorem describes structurally stable planar systems. This theorem states that they are structurally stable if and only if:
\begin{romanlist}[(iii)]
\item{all its critical points are hyperbolic (i.e., stable, unstable or saddles)}
\item{all periodic orbits have a multiplier which is different from 1}
\item{there is not phase curves which connect saddles.}
\end{romanlist}
In the special case of subclass of cosmological models which evolution is described by some vector field $[f, g]^T$ on the Poincar{\'e} sphere, where $f$ and $g$ are polynomials of degree $m$. There is a simple method of detection structurally stable (and generic) dynamical systems. Vector field $f$ is structurally stable if
\begin{romanlist}[(iii)]
\item{the number of critical points and limit cycles is finite}
\item{there is no trajectories connecting saddle points.}
\end{romanlist}
Note that the presence of non-hyperbolic critical points in the phase portrait is a sufficient condition for structural instability. It can be also shown that if the polynomial vector field is structurally stable on the Poincar{\'e} sphere $S^2$, then the corresponding vector field $\mathbf{X}=[f,g]^T$ is structurally stable on $R^2$ \cite{Perko:2001}.

Following the Peixoto theorem the structural stability is a generic property of the $C^1$ vector field on a compact two-dimensional differentiable manifold $M$.

From the mathematical point of view the typical (generic) cases are more interesting than non-typical ones because the latter are more complicated which makes it difficult to analyse. This prejudice is shared by all dynamists \cite{Abraham:book}. All the effective and reliable models should possess some kind of a stability. Otherwise we have situation that models may be fragile and many dramatically different models become in good agreements with observation. Rene Thom's opinion is that such a degeneration would be fatal for modelling in science. In some sense, the structural stability of the model explains why simple models describe a complex universe and can be in agreement agreeing
with observations.

\subsection{Road toward structurally stable cosmological models}

Let us discuss the phase portraits of flat cosmological systems with the scalar field and matter in details. The phase variables are chosen as energetic variables \cite{Szydlowski:2008in}. In this Section we will consider different cases of the models with the canonical and phantom scalar field with a quadratic potential.

From the point of view of structural stability the $\Lambda$CDM model and its counterpart in scalar field cosmology are structurally stable. We are looking for SFC models with such a form of the potential which makes the system structurally stable.

\begin{figure}
\begin{center}
\includegraphics[scale=0.4]{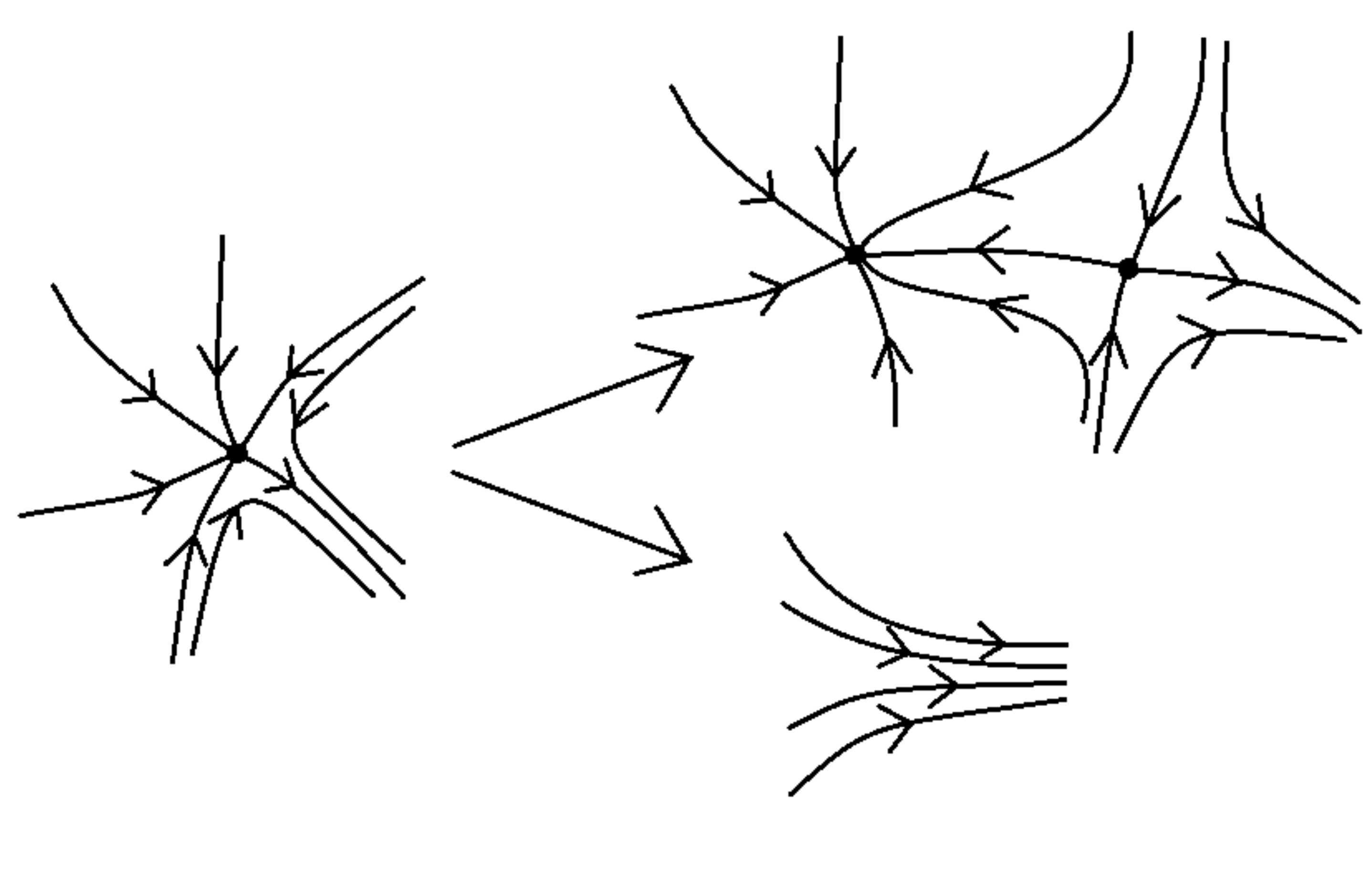}
\caption{One way toward structurally stable cosmological system. Different scenarios of a transition from the structurally unstable critical point to the structurally stable one via the saddle node bifurcation.}
\label{fig:2}
\end{center}
\end{figure}

The characteristic scenario of the saddle-node bifurcation is the following (see figure \ref{fig:2}):
\begin{romanlist}[(iii)]
\item{a saddle-node critical point located at the circle at infinity bifurcates to the saddle and node separately}
\item{a critical point under small perturbation vanishes and does not appear on the phase portrait.}
\end{romanlist}

\begin{figure}
\begin{center}
\includegraphics[scale=0.55]{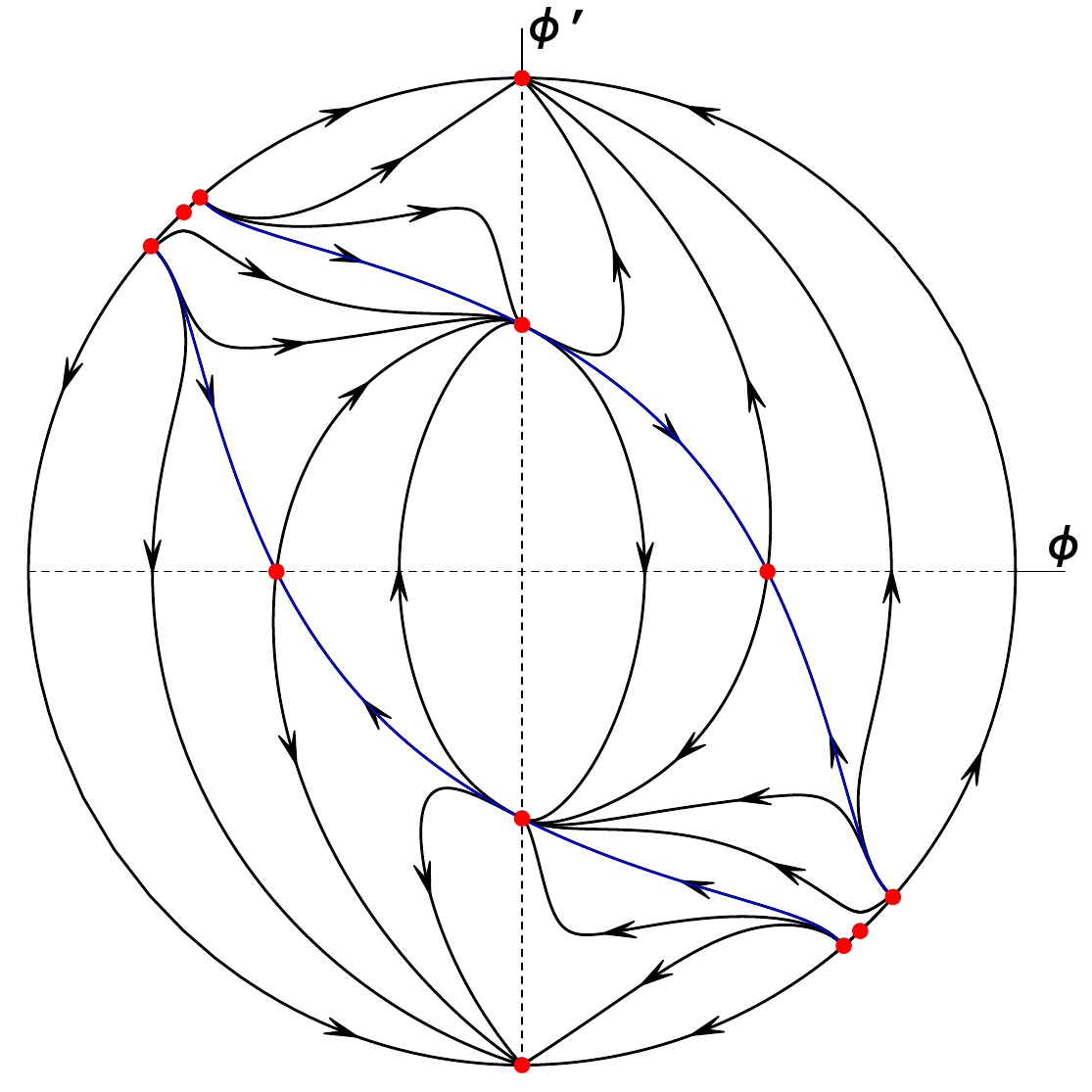}
\includegraphics[scale=0.55]{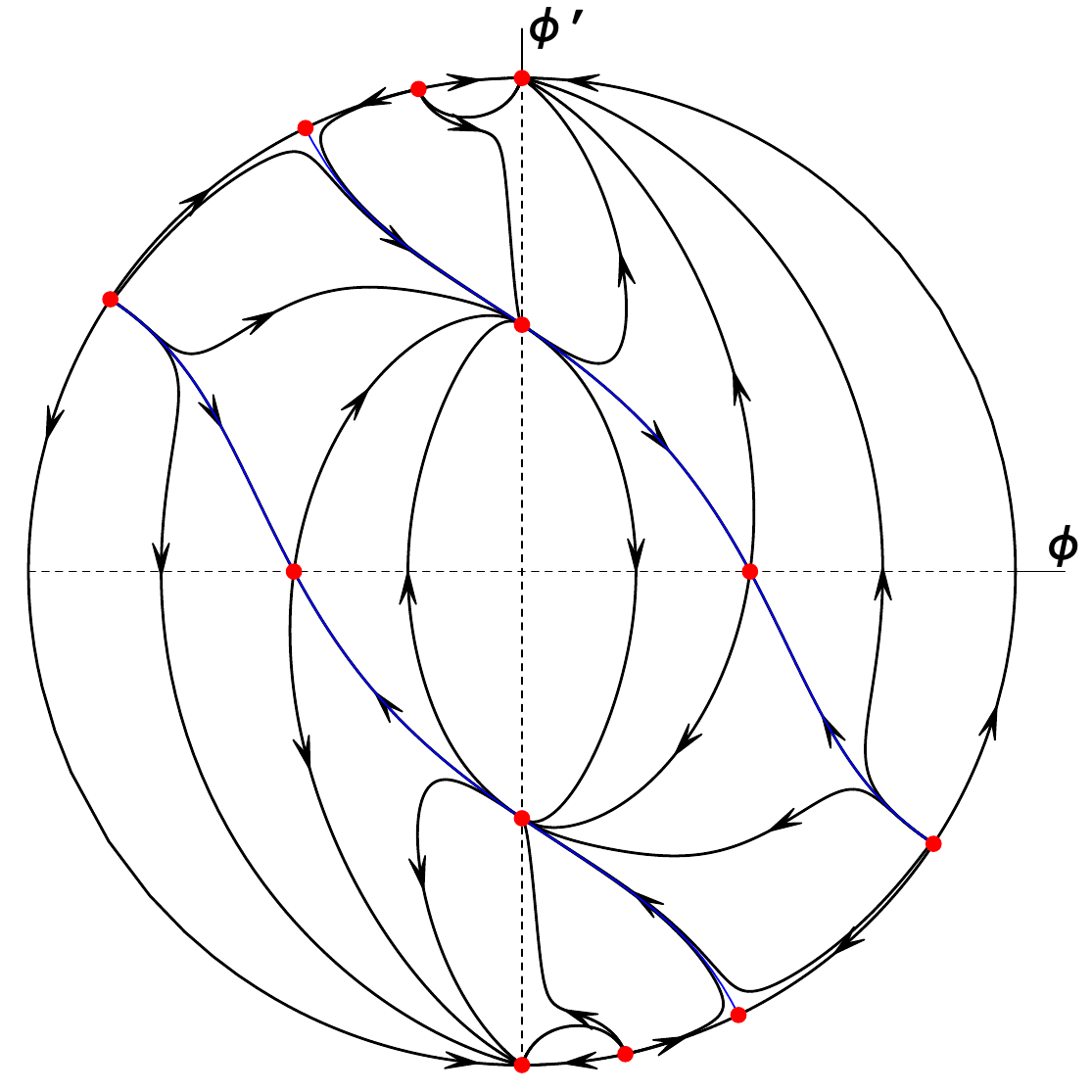}
\caption{The phase space portraits for the canonical scalar field and coupling constant $\frac{1}{6}<\xi<\frac{3}{16}$ (left) and $\frac{3}{16}<\xi<\frac{1}{4}$ (right). The systems are structurally unstable because of the presence of the saddle-node critical point at the circle at infinity.}
\label{fig:3}
\end{center}
\end{figure}

\begin{figure}
\begin{center}
\includegraphics[scale=0.65]{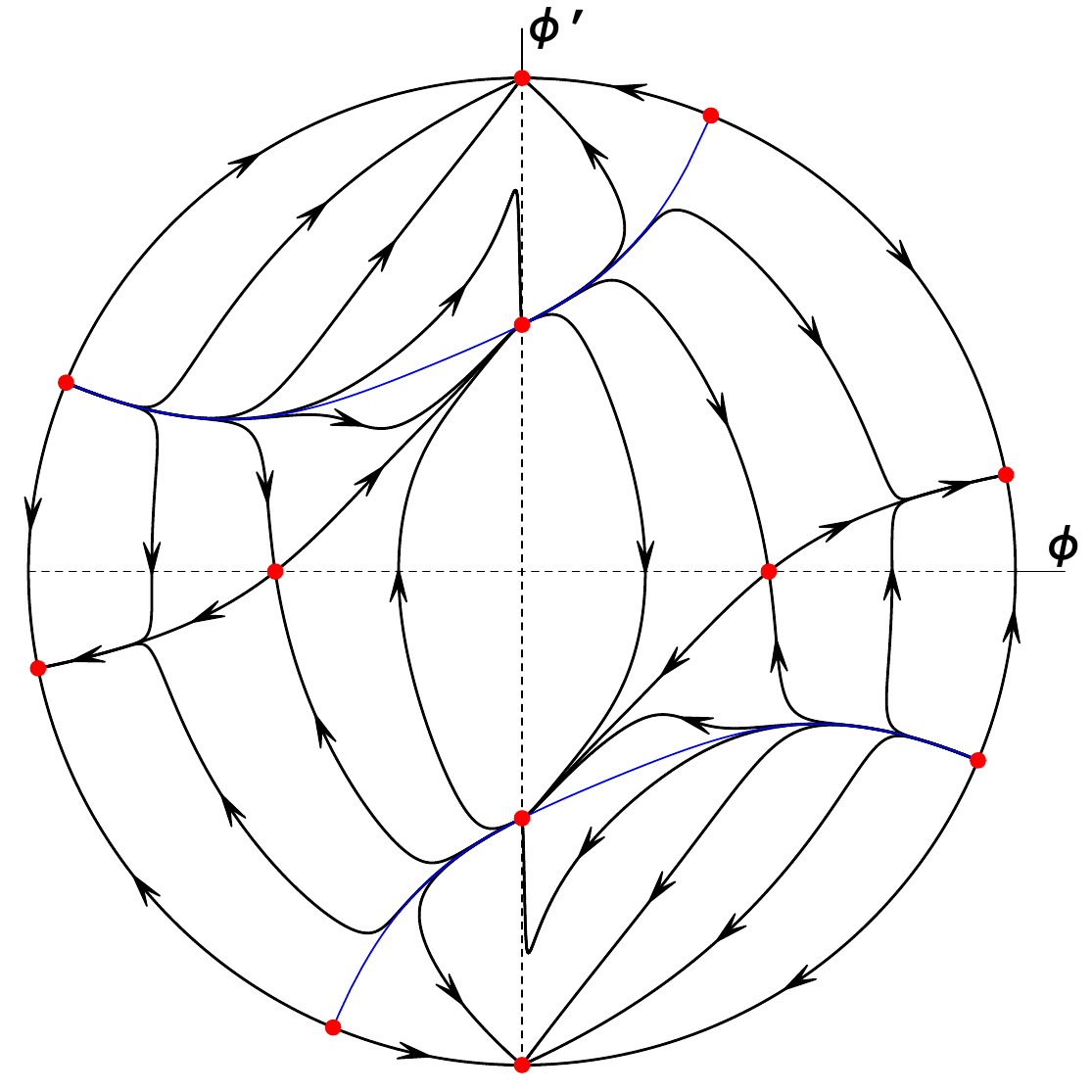}
\caption{The phase portrait for the canonical scalar field for a coupling constant $\xi<0$. The system is structurally unstable because of the presence of the saddle-node critical point at the circle at infinity.}
\label{fig:4}
\end{center}
\end{figure}

\begin{figure}
\begin{center}
\includegraphics[scale=0.55]{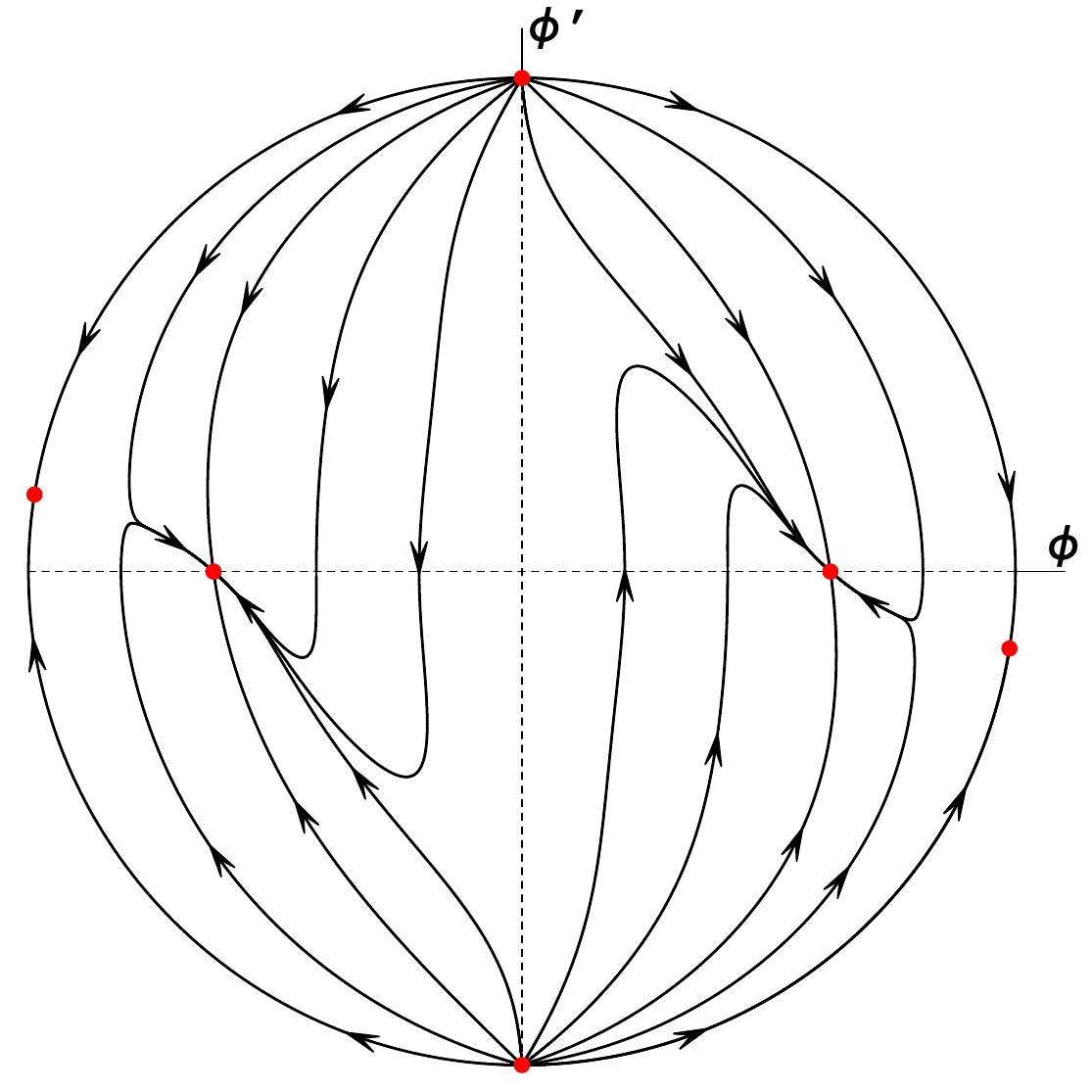}
\includegraphics[scale=0.55]{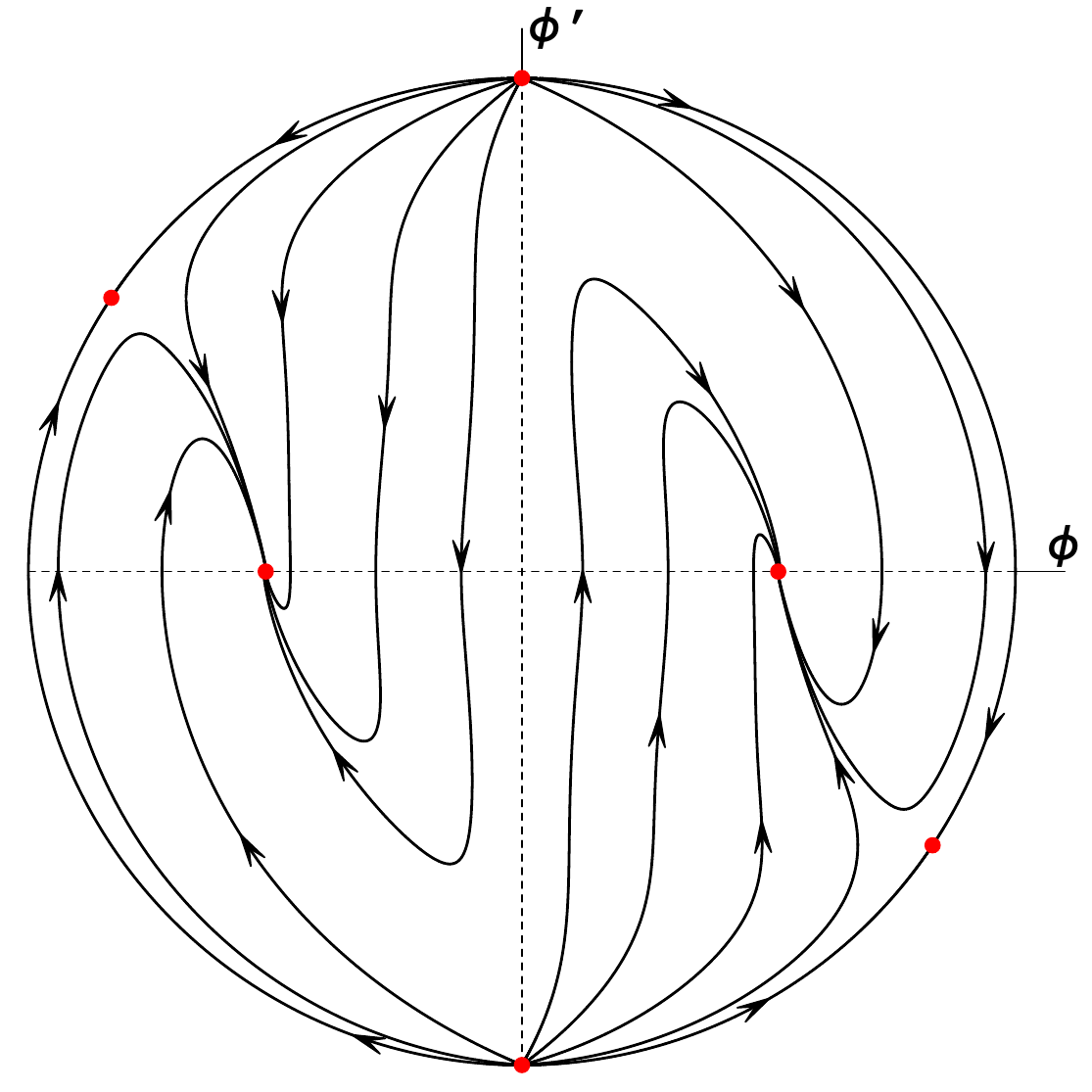}
\caption{The phase portrait for the phantom scalar field for a coupling constant $0<\xi<\frac{3}{25}$ (left) and $\frac{3}{25}<\xi<\frac{1}{6}$ (right). The system is structurally unstable because of the presence of the saddle-node critical point at the circle at infinity.}
\label{fig:5}
\end{center}
\end{figure}

\begin{figure}
\begin{center}
\includegraphics[scale=0.55]{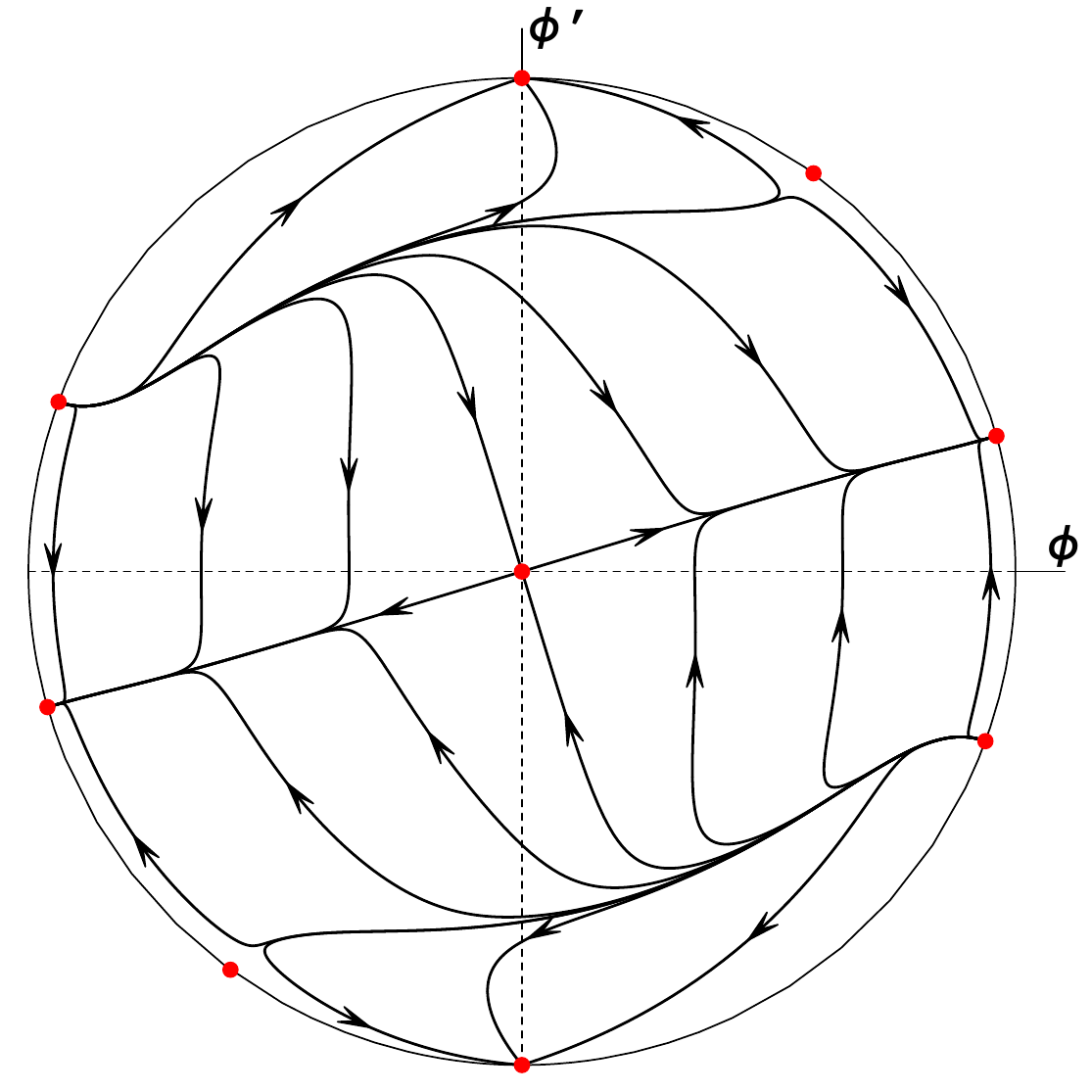}
\includegraphics[scale=0.55]{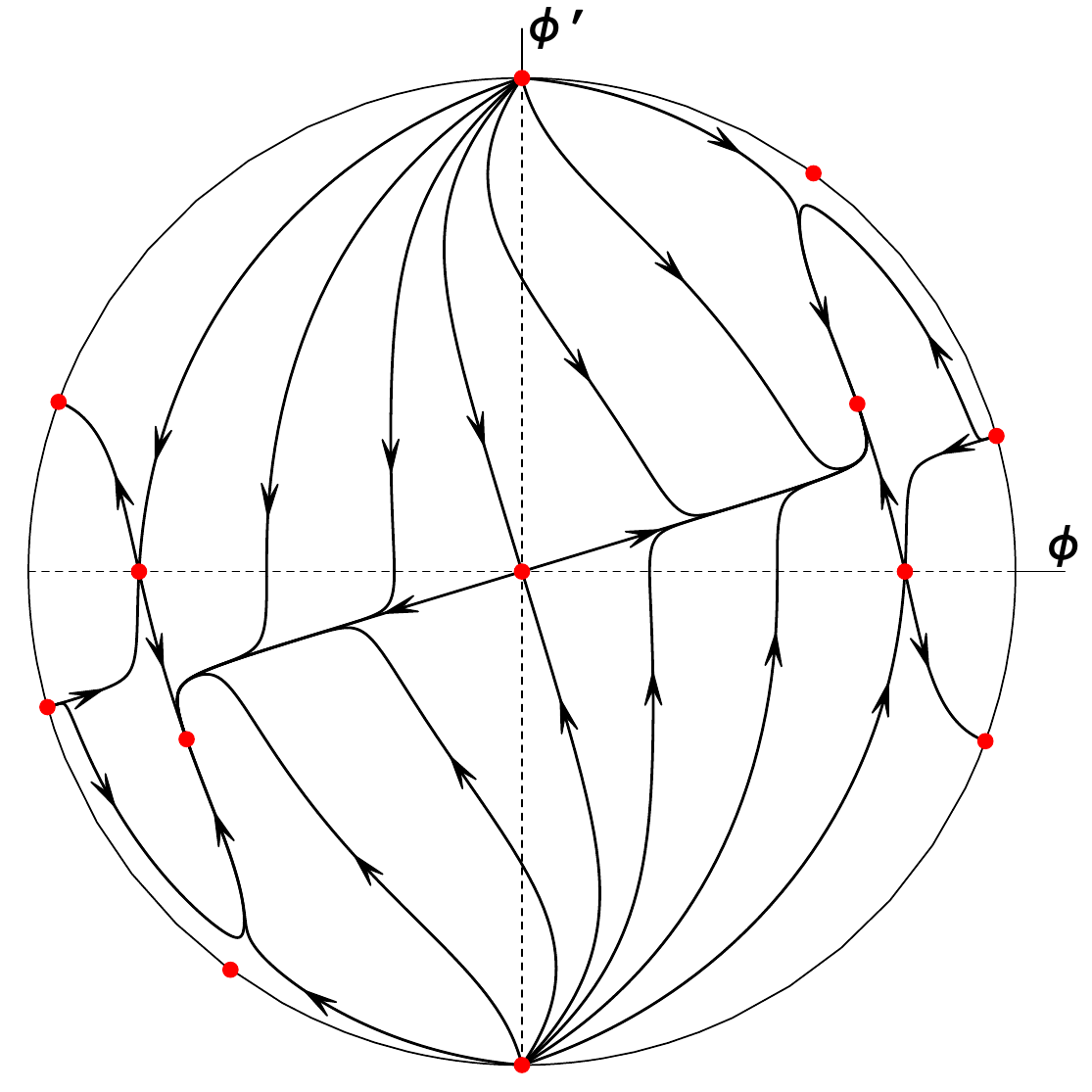}
\caption{The phase space portrait for the non-minimally coupled canonical scalar field $\epsilon=1$ (left) and phantom scalar field $\epsilon=-1$ (right) with $\xi=-\frac{1}{12}$ and constant potential function. All the critical points are hyperbolic and the system is structurally stable.}
\label{fig:6}
\end{center}
\end{figure}

\begin{figure}
\begin{center}
\includegraphics[scale=0.55]{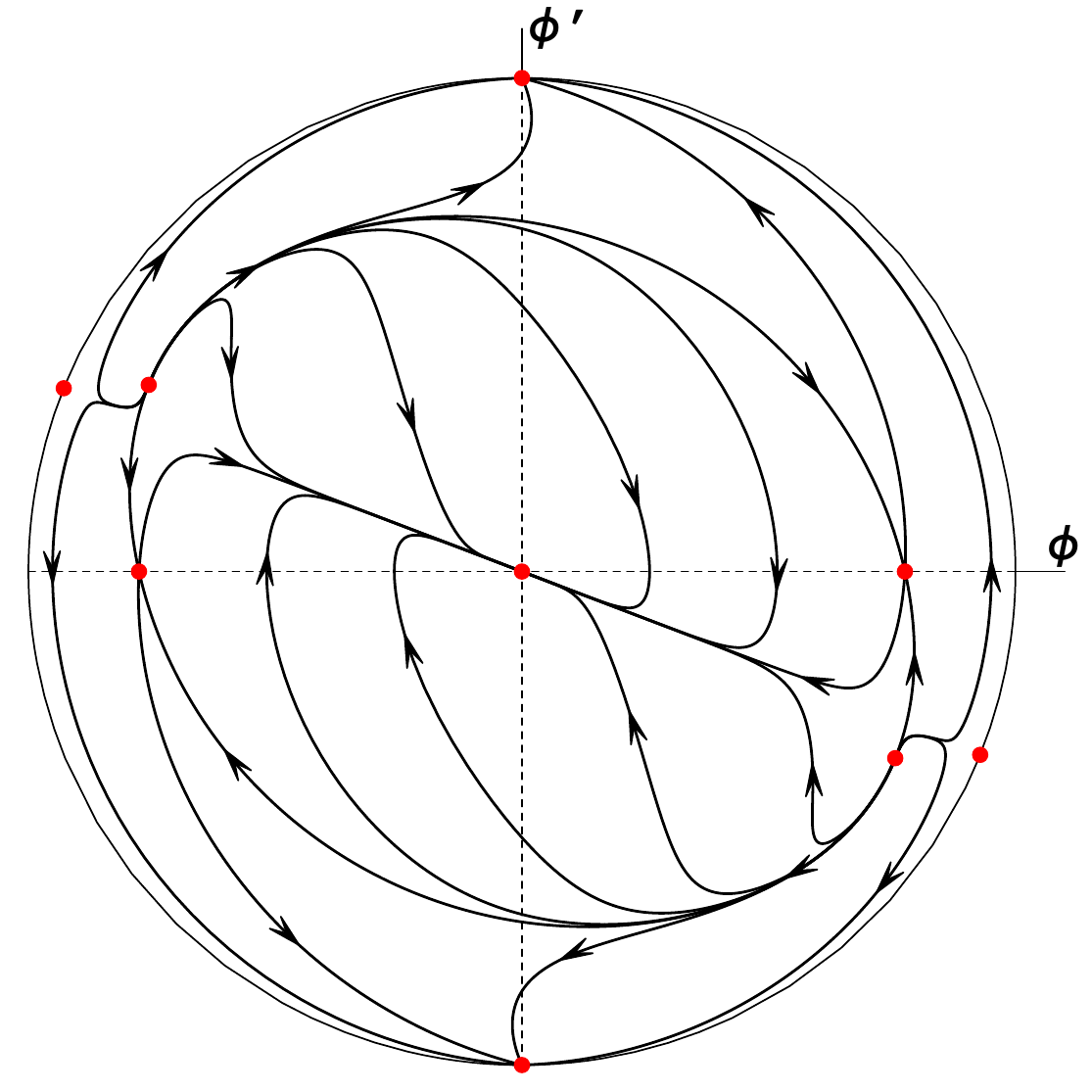}
\includegraphics[scale=0.55]{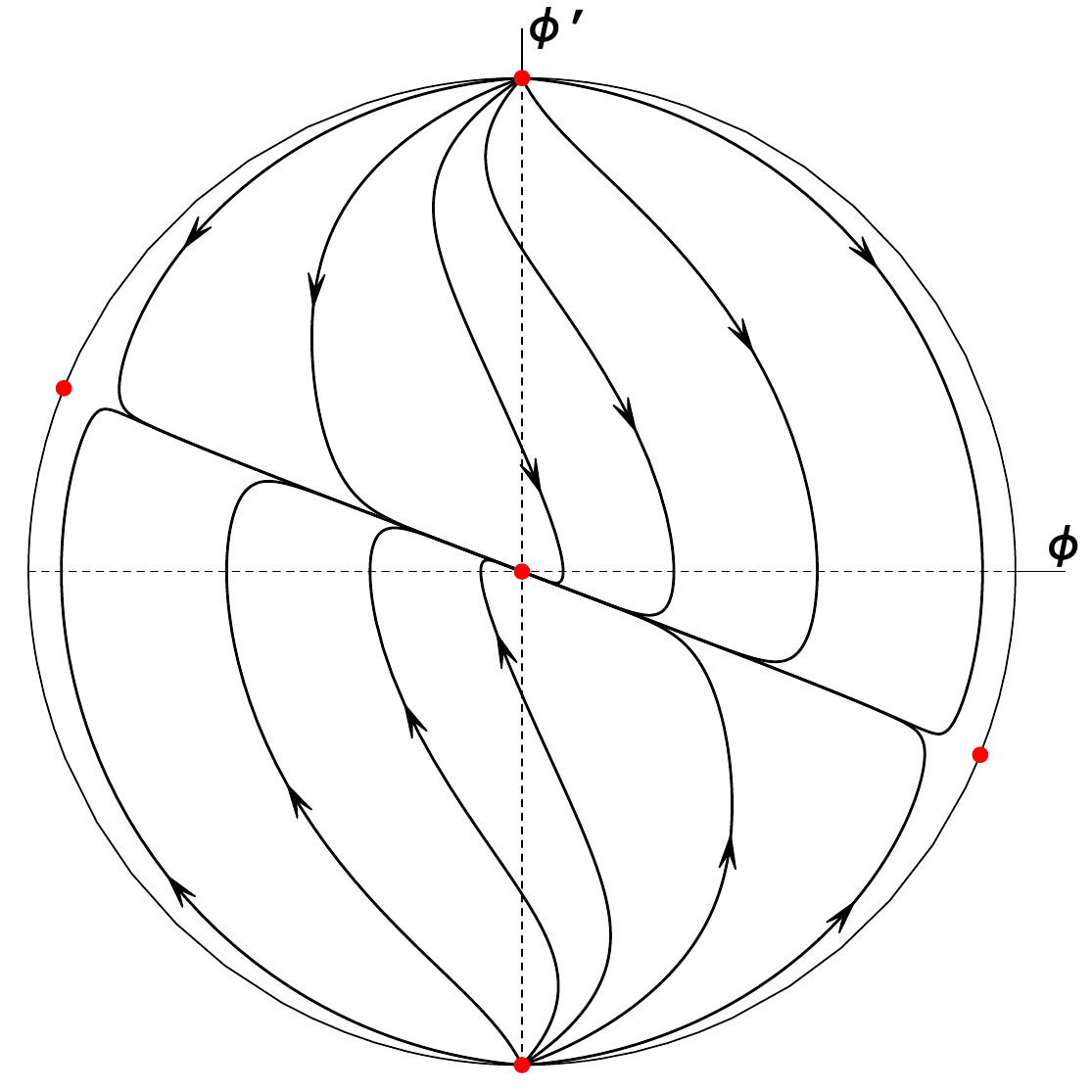}
\caption{The phase space portrait for the non-minimally coupled canonical scalar field $\epsilon=1$ (left) and phantom scalar field $\epsilon=1$ (right) with $\xi=\frac{1}{12}$ and constant potential function. All the critical points are hyperbolic and the system is structurally stable.}
\label{fig:7}
\end{center}
\end{figure}

\begin{figure}
\begin{center}
\includegraphics[scale=0.55]{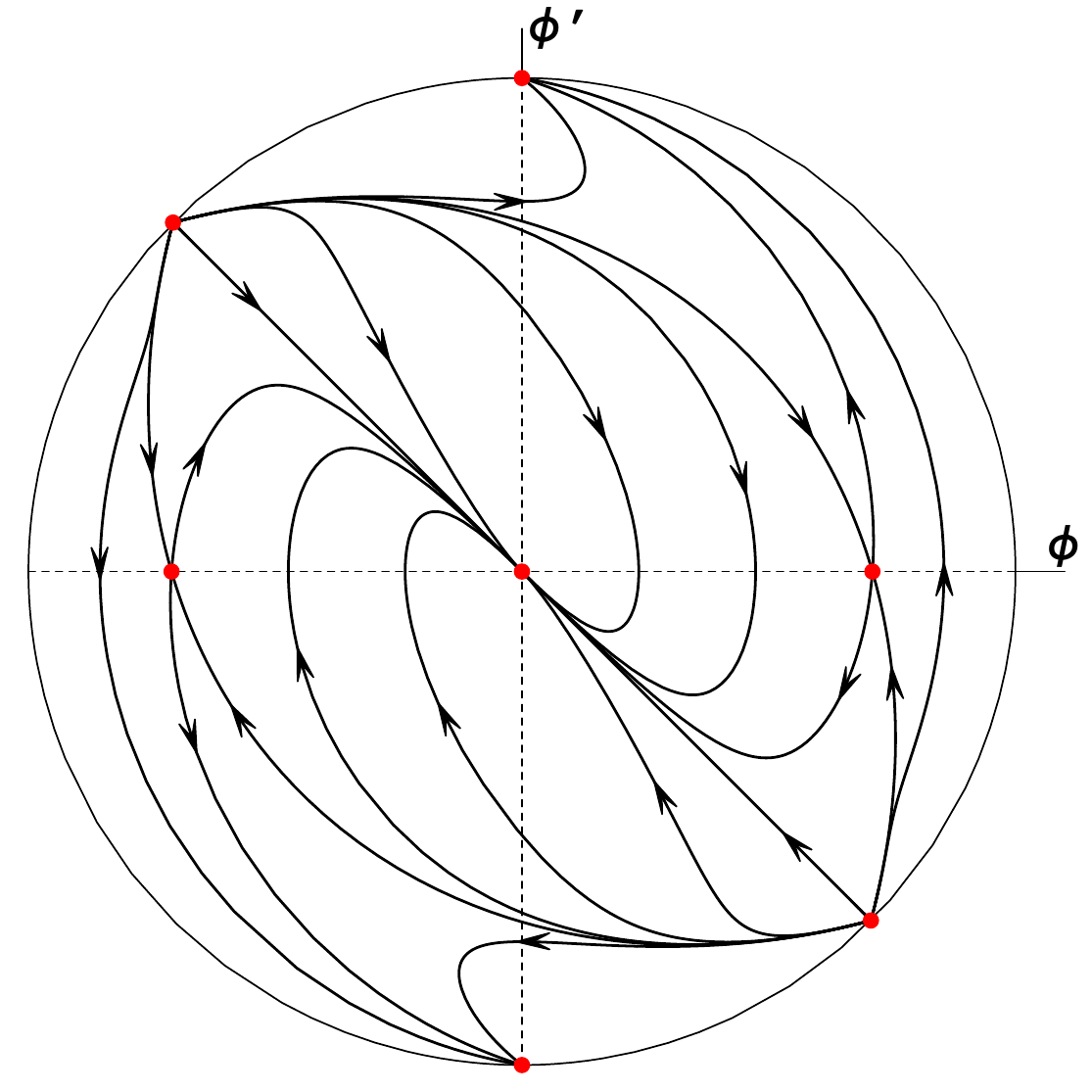}
\includegraphics[scale=0.55]{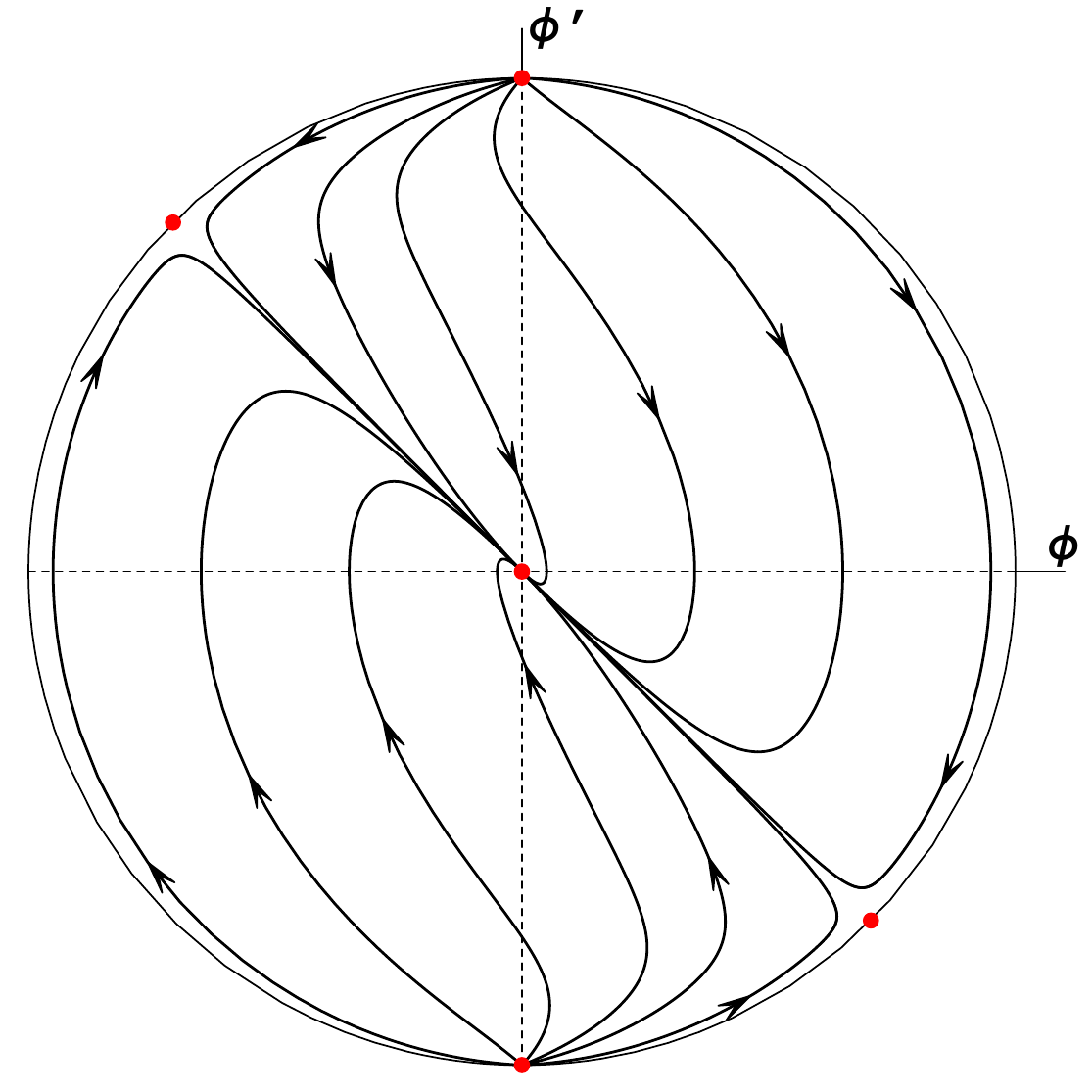}
\caption{The phase space portraits for the conformally compled $\xi=\frac{1}{6}$ canonical $\epsilon=1$ (left) and phantom $\epsilon=-1$ (right) scalar fields and constant potential function. All the critical points are hyperbolic and the system is structurally stable.}
\label{fig:8}
\end{center}
\end{figure}

In figures \ref{fig:3}, \ref{fig:4}, \ref{fig:5}, \ref{fig:6}, \ref{fig:7} and \ref{fig:8}, zoo cosmological models of the SFC with the non-minimal coupling constant is presented on the plane phase.
We observe that phase portraits change dramatically under changing the form of potential function : figures \ref{fig:3}, \ref{fig:4} and \ref{fig:5} represent phase space diagrams for a quadratic potential function while  figures \ref{fig:6}, \ref{fig:7} and \ref{fig:8} represent phase space diagrams for a constant potential function. They are very sensitive to the choice of the potential form. 

The phase portraits illustrate how the SFC models make structurally stable systems by introduction some perturbation in the form of the non-minimal coupling constant for suitable potential form. Different phase portraits illustrate a road toward structurally stable models via non-minimal coupling constant effect.

It is characteristic in many examples of structurally unstable models that a road toward structurally stable ones is thoroughly saddle-node bifurcation (figure \ref{fig:2}).

All phase portraits on figures \ref{fig:3}, \ref{fig:4} and \ref{fig:5} represent structurally unstable critical points. They represent cosmological evolution with matter in the form of canonical scalar field non-minimally coupled to gravity with a quadratic form of potential.
The structural instability is an argument for us, that such a form of a potential, although very popular, is not physically realistic. We are looking for such a form of a potential which gives rise to the structurally stable model which form open and dense subsets in the ensemble. Such subsets form SFC models with the constant potential (see figures \ref{fig:6}, \ref{fig:7} and \ref{fig:8}).

\subsection{Structurally stable cosmological models which reproduce an expansion of the Universe}

It is well know that the $\Lambda$CDM model is structurally stable \cite{Szydlowski:2012zz}. This fact explains a property of flexibility of this model with respect to the astronomical data. The $\Lambda$CDM model is the best candidate for an explanation of the accelerating phase of the expansion of the current Universe. This conclusion follows from the Bayesian analysis of a sample of theoretical models used for an explanation of the acceleration \cite{Capozziello:2009te,Bamba:2012cp,Szydlowski:2006ay}.

It is interesting to answer the question what a class of models of SFC possesses also the property of structural stability.

To make both class of models to be commensurable, we assume that the scalar field as well as the potential depend on cosmic time through the scale factor, i.e.
\begin{equation}
    \rho_{\phi}=\frac{1}{2}\left(\frac{\ud \phi}{\ud a}
\right)^2 H^2(a) a^2+V(\phi(a))\,,
\end{equation}
\begin{equation}
    p_{\phi}=\frac{1}{2}\left(\frac{\ud \phi}{\ud a}\right)^2 H^2(a) a^2-V(\phi(a))\,,
\end{equation}
where $H(a)$ is the Hubble function. We compare two models, the model with the canonical scalar field and a dust matter for which $H(t)$ satisfies the following relation
\begin{equation}
    \frac{\ud H}{\ud t}=-\frac{1}{2}\big(\rho_\phi+\rho_{m,0}a^{-3}+p_\phi+0\big)\,,
\end{equation}
with the standard cosmological model for which
\begin{equation}
\label{eq:Hlcdm}
    H^2(a)=H_0^2(\bar{\Omega}_{m,0}a^{-3}+(1-\bar{\Omega}_{m,0})).
\end{equation}
Then from the acceleration equation
\begin{equation}
\label{eq:phia}
    \left(\frac{\ud\phi}{\ud a}\right)^{2}=-\frac{a\frac{\ud H^2(a)}{\ud a}+\rho_{m,0}a^{-3}}{a^{2}H^2(a)},
\end{equation}
after substitution $H^2(a)$ from \eqref{eq:Hlcdm} to \eqref{eq:phia} , we obtain
\begin{equation}   
\left(\frac{\ud\phi}{\ud z}\right)^2=\frac{(1+z)\big(3\bar{\Omega}_{m,0}-\Omega_{m,0}\big)}{\bar{\Omega}_{m,0}(1+z)^3+(1-\bar{\Omega}_{m,0})},
\end{equation}
where $z$ is redshift $1+z=a^{-1}$ and $\Omega_i\equiv\frac{\rho_i}{3H_0^2}$ are dimensionless density parameters and $H_0$ is the present value of the Hubble function. Above formula reduces to the integral
\begin{equation}
\phi(z)=\frac{\sqrt{3\bar\Omega_{m,0}-\Omega_{m,0}}}{\sqrt{\bar\Omega_{m,0}}}\int_1^{1+z}\sqrt{\frac{u}{u^3+\alpha}}\ud u,
\end{equation}
where $u\equiv(1+z)$ and $\alpha\equiv\frac{1-\bar{\Omega}_{m,0}}{\bar{\Omega}_{m,0}}$.
The integration gives
\begin{equation}
\label{eq:phisol_1}
    (1+z)^{3/2}+\sqrt{\alpha+(1+z)^3}=\phi_0 e^{\lambda\phi},
\end{equation}
where $\phi_0=1+\sqrt{\alpha+1}$, $\lambda_0\equiv\frac{3}{2}\sqrt{\frac{\bar{\Omega}_{m,0}}{3(\bar\Omega_{m,0}-\Omega_{m,0})}}$.
Relation \eqref{eq:phisol_1} can be rewritten to the form
\begin{equation}
    \sqrt{U}+\sqrt{U+\alpha}=\phi_0 e^{\lambda\phi}
\end{equation}
or
\begin{equation}
    \sqrt{U}-\sqrt{U+\alpha}=-\alpha \phi_0^{-1} e^{-\lambda\phi}.
\end{equation}
From these equations one can derive the exact formula for $U(\phi)$
\begin{equation}
    U(\phi)=\frac{1}{4}\left(\phi_0 e^{\lambda \phi}-\frac{\alpha}{\phi_0} e^{-\lambda \phi}\right)^2
\end{equation}
or
\begin{equation}
\label{eq:z_of_phi}
    (1+z)^3=\frac{1}{4}\left(\phi_0 e^{\lambda \phi}-\frac{\alpha}{\phi_0}e^{-\lambda\phi}\right)^2
\end{equation}
describing the evolution of the scalar field.

From the relation \eqref{eq:z_of_phi} we obtain an equivalent relation between the scalar field $\phi$ and the scale factor
\begin{equation}
    a(\phi)=\frac{\sqrt[3]{4}}{\Big(\phi_0 e^{\lambda \phi}-\big(\frac{\alpha}{\phi_0}\big)e^{-\lambda\phi}\Big)^{\frac{2}{3}}}.
\end{equation}
For large $z$ (or small values of the scale factor) we obtain asymptotic expressions
\begin{equation}
    z=\left(\frac{\phi_0}{2}\right)^{\frac{2}{3}}e^{\frac{2}{3}\lambda\phi}
\end{equation}
or
\begin{equation}
    a\propto e^{-\frac{2\lambda}{3}\phi}.
\end{equation}
Note that if we choose the $\alpha=\phi_0^2$, then function $U(\phi)$ is of the $\sinh^2(\phi)$ type.
The potential of the scalar field finally takes the following form
\begin{equation}
    V(\phi)=\frac{3H_0^2}{2}\left(\left(\bar\Omega_{m,0}-\Omega_{m,0}\right)\left(\frac{1}{4}\right)\left(\phi_0 e^{\lambda \phi}-\frac{\alpha}{\phi_0}e^{-\lambda\phi}\right)^2+2\left(1-\bar\Omega_{m,0}\right)\right).
\end{equation}
As a result we have found a class of scalar field cosmological models which reproduces the $\Lambda$CDM model. Note that, if we put the constant $\alpha=0$, then we obtain the potential of the form $V_0+\beta e^{\lambda\phi(a)}$, and $\phi(a)=-\frac{3}{2\lambda}\ln\frac{a}{a_0}$. Therefore the models with this potential will be structurally stable automatically.

If we considered the distance between the $\Lambda$CDM model and the model mimicking this model in SFC then it is zero in the sense of the metric \eqref{eq:met_d}. However, the cosmological constant possesses a simple interpretation in the SFC and appears as an effective and an emergent parameter from the scalar field potential.

\section{Symmetry group of cosmological dynamical systems}

When considering cosmological models a symmetry group as dynamical systems can be useful in searching for the exact solution of the equation. Investigations of the Lie symmetries of cosmological models represented by the Lagrange function have recently been very popular. Such symmetries are called Noether symmetries \cite{Capozziello:2009te, Hydon:book, Olver:book, Stephani:book}.

Let us consider dynamics of cosmological models as a system of differential equations (non-autonomous in general)
\begin{equation}
\label{eq:sys_gen}
    \frac{\ud u^\alpha}{\ud t}=f^\alpha(t,u), \quad \alpha=1,...,n, \quad u=(u^1,...,u^n),
\end{equation}
where $t \in R$ and $u^\alpha\in R^n$ and there are point-point transformations
\begin{equation}
T \colon u^\alpha\longmapsto\bar{u}^\alpha=\bar{u}^\alpha(t,u), \quad
t\longmapsto\bar{t}=\bar{t}(t,u),
\end{equation}
which map each solution of the system \eqref{eq:sys_gen} into a solution of the same system. The transformation $T$ is a Lie group and it is called Lie symmetries of differential equations.

Let $M$: $u^\alpha=u^\alpha(t)$ be a solution of \eqref{eq:sys_gen}. $M$ is a submanifold in $R^n$. If $\bar{M}=T(M)$ then we have
\begin{equation}
    \bar{M} \bar{t}=\bar{t}(t,u(t))
\bar{u}^\alpha=\bar{u}^\alpha(t,u(t)).
\end{equation}

The derivatives $p^\alpha=\frac{\partial u^\alpha}{\partial t}$, $\bar{p}^\alpha=\frac{\partial \bar u^\alpha}{\partial \bar{t}}$ satisfy the condition
\begin{equation}
    \bar{p}^\alpha D\bar{t}=D\bar{u}^\alpha,
\end{equation}
where $D=\frac{\partial}{\partial t}+p^\alpha \frac{\partial}{\partial u^\alpha}$. By solving the above equation we obtain a relation $\bar{p}^\alpha=\bar{p}^\alpha(t,u,p)$. After adjoining these solutions to the finite transformations $T$, one gets a new set of transformations $\bar T$ which are called an extension of $T$ transformation on the first derivatives.

If the infinitesimal operator of the group $T$ has the form
\begin{equation}
    X=\xi(t,u)\frac{\partial}{\partial t}+\eta^\alpha(t,u)\frac{\partial}{\partial u^\alpha}
\end{equation}
then the prolongation $X$ on the group $\tilde T$ on the first derivatives has a form
\begin{equation}
\label{eq:62}
    \tilde X = X+(D\eta^\alpha-p^\alpha D\xi)\frac{\partial}{\partial p^\alpha}.
\end{equation}

The functions $\mathcal F(t,u,p)$ defined on $R^{1+n+n}$ (the space of state variables, its derivatives and time) which are preserved under transformations $\tilde T$ are called invariants of the group $T$ and may be useful in choosing new variables convenient for the integration of the system. They satisfy equation
\begin{equation}
\label{eq:63}
    \tilde{X}(\mathcal{F})=0
\end{equation}
and vice versa, functions which satisfy \eqref{eq:63} are invariants of the group generated by $X$. The applications equation \eqref{eq:63} to the original equation \eqref{eq:sys_gen} gives
\begin{equation}
\label{eq:64}
    \tilde{X}(p^\alpha-f^\alpha(t,u))=0
\end{equation}
Inserting \eqref{eq:62} into \eqref{eq:64} and taking into account $p^\alpha=f^\alpha(t,u)$, we get a system of partial differential equations which give necessary and sufficient conditions for admissible symmetries of differential equations
\begin{equation}
    \frac{\partial \eta^\alpha}{\partial t}+\frac{\partial \eta^\alpha}{\partial u^\beta}f^\beta-\frac{\partial \xi}{\partial t}f^\alpha-\frac{\partial\xi}{\partial u^\beta}f^\beta f^\alpha=\xi\frac{\partial f^\alpha}{\partial t}+\eta^\beta\frac{\partial f^\alpha}{\partial u^\beta};\ \alpha,\ \beta=1,...,n.
\end{equation}
In the following we shall consider an autonomous dynamical system of a cosmological origin for which we have $\frac{\partial f^\alpha}{\partial t}=0$. We assume that $\xi(t, u)= \xi(t)$ and $\eta^\alpha(t,u)=\eta^\alpha(u)$, i.e. that we restrict our study to the so called quasihomological symmetries. In the special case if $\xi(t)\propto t$ (some reasons of such a choice can be found in \cite{Aguirregabiria:2003uh}) then we obtain a case of so-called homological transformations. They are important in searching for so-called scaling solutions. In our terminology by the scaling type of solutions we understand such trajectories in the phase space which admit homological symmetry transformations. If we considered the quasi-homological transformation generated by an infinitesimal operator $X=\xi(t)\frac{\partial}{\partial t}+\eta^{\beta} (u)\frac{\partial}{\partial u^\beta}$, then conditions of admissibility of such symmetries simplify to the form
\begin{equation}
    \frac{\partial \eta^\alpha}{\partial u^\beta}f^\beta-\frac{\partial \xi}{\partial t}f^\alpha=\eta^\beta\frac{\partial f^\alpha}{\partial u^\beta}.
\end{equation}
The results of an application of the Lie symmetry of a group analysis to the case of a simple SFC model are summarised in the following theorem.

Let $V=V(\phi)$ a potential of the scalar field be reversible function $\phi^{-1}=\phi^{-1}(V)$ in some interval. Let us consider SFC given in the form of dynamical system describing a model of a cosmological evolution with a single canonical scalar field, the potential $V(\phi)$ and a barotropic matter, then
\begin{equation}
\label{eq:dyn_exp}
\begin{split}
    x' &= -3x-\frac{1}{\sqrt 6}w^2 W(x,\ z,\ w)+x\left(3x^2+\frac{3}{2}\gamma z^2\right), \\
    z' &= z\left(3x^2+\frac{3}{2}\gamma z^2-\frac{3}{2}\gamma\right), \\
    w' &= w\left(3x^2+\frac{3}{2}\gamma z^2\right),
\end{split}
\end{equation}
where $x=\frac{\dot\phi}{\sqrt 6 H}$, $z=\frac{\sqrt \rho}{\sqrt 3 H}$, $w=\frac{1}{H}$, $\gamma$ is a parameter from state equation $p=(\gamma-1)\rho$, $V(x,\ w,\ z)=\frac{3}{w^2}(1-x^2-z^2)$ and $U(\phi)=\frac{\partial V(\phi)}{\partial \phi}=W(V(\phi))=W(V(x,w,z))$.

If the symmetry operator has the form $X=h t \frac{\partial}{\partial t}+\eta^1 (x) \frac{\partial}{\partial x}+\eta^2 (w) \frac{\partial}{\partial w}+\eta^3 (z) \frac{\partial}{\partial z}$ and $\eta^1(x)=ax,\ \eta^2(w)=bw,\ \eta^3(z)=cz$, then the conditions of the admissibility of the homological transformation has the form
\begin{equation}
\label{eq:hom_sys}
\begin{split}
    & 9x(-2h+2(2a+h)x^2+(2 c + h) z^2 \gamma) +
  \sqrt{6} ((a - 2 b - h) w^2 W(V)+\\  
    & + 6 (a x^2 + c z^2 - b (-1 + x^2 + z^2)) \frac{\partial W(V)}{\partial V}) = 0\,,\\
    & z(2 (2 a - b + c + d) x^2 + \gamma (b - c - d + (-b + 3 c + d) z^2))=0\,,\\
    & w (2 (2 a + b - c + h) x^2 + (b + c + h) z^2 \gamma) = 0.
\end{split}
\end{equation}
These homological transformations forced the following form of the potential in a codimension 2 domain $\big\{$ $(x,\ z,\ w )$:  \eqref{eq:dyn_exp} and \eqref{eq:hom_sys} are satisfied $\big\}$
\begin{equation}
    V(\phi)=V_0 e^{\lambda\phi} \quad \text{or} \quad V(\phi)= \Lambda.
\end{equation}
This relation is also preserved if instead of homological transformation quasi-holomological symmetry is postulated.

The phase space of the scalar field cosmological model with the matter is considered in the paper \cite{Copeland:1997et} in the context of existence scaling solutions. Authors proved that such solution exist for an exponential potential of the scalar field as an attractor in 2D. Scaling solutions for the potential $V(\phi)=V_0 e^{\lambda\phi}$ have following equations: $x=x_0$, $z=z_0$ and $w=w(\tau)$ is an arbitrary function.

\begin{figure}
\begin{center}
\includegraphics[scale=0.45]{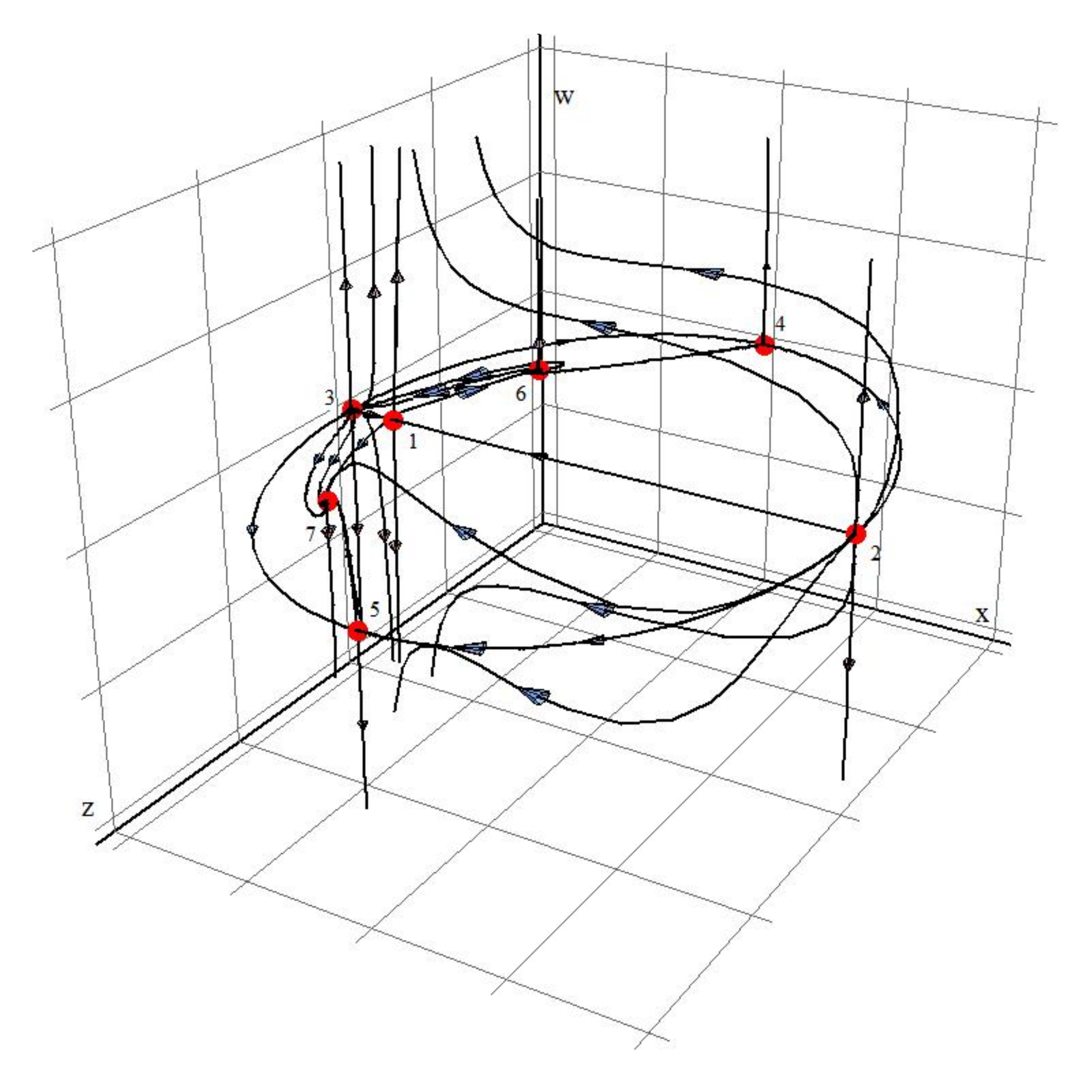}
\caption{A phase portrait for cosmological model (equations \eqref{eq:dyn_exp}) with an exponential potential (for example we put $\lambda=2$) and dust matter. Critical points (1), (4) and (5) are of saddle types. Critical points are (2) and (3) are of unstable node types. Critical points (6) and (7) are of saddle types. Invariant submanifold ${w=0}$ is stable submanifold and trajectories and separatrices starting from critical points (6) and (7) are one dimensional unstable submanifolds.  The critical points (6) 
and (7) are structural stable. The critical points (6) and (7) are representing scaling solutions.}
\label{fig:9}
\end{center}
\end{figure}

The critical points of the system \eqref{eq:dyn_exp} and an exponential form of the potential 
function forced by homology \eqref{eq:hom_sys} are completed in Table \ref{tab:3} while the phase portrait for this system is presented in Fig. \ref{fig:9}.

\begin{table}
\tbl{Critical points for the dynamic system \eqref{eq:dyn_exp} with an exponential potential function.\label{tab:3}}
{\begin{tabular}{ccccc} \toprule
   No & $x$ & $z$ & $w$ & dominating form of energy \\ \colrule
$1$ & $-\sqrt{\frac{\lambda}{6}}$ & $0$ & $0$  & scalar field\\ 
$2$ & $1$ & $0$ & $0$  & kinetic part of scalar field\\
$3$ & $-1$ & $0$ & $0$  & kinetic part of scalar field\\
$4$ & $0$ & $1$ & $0$  &  barotropic matter\\
$5$ & $0$ & $-1$ & $0$   & barotropic matter\\
$6$ & $-\sqrt{\frac{3}{2}}\frac{\gamma}{ \lambda}$ & $\sqrt{-\frac{3\gamma}{ \lambda^2}+1}$ & $0$   & scalar field and barotropic matter\\
$7$ & $-\sqrt{\frac{3}{2}}\frac{\gamma}{ \lambda}$ & $-\sqrt{-\frac{3\gamma}{ \lambda^2}+1}$ & $0$ & scalar field and barotropic matter\\  \botrule
\end{tabular}}
\end{table}

\begin{figure}
\begin{center}
\includegraphics[scale=0.45]{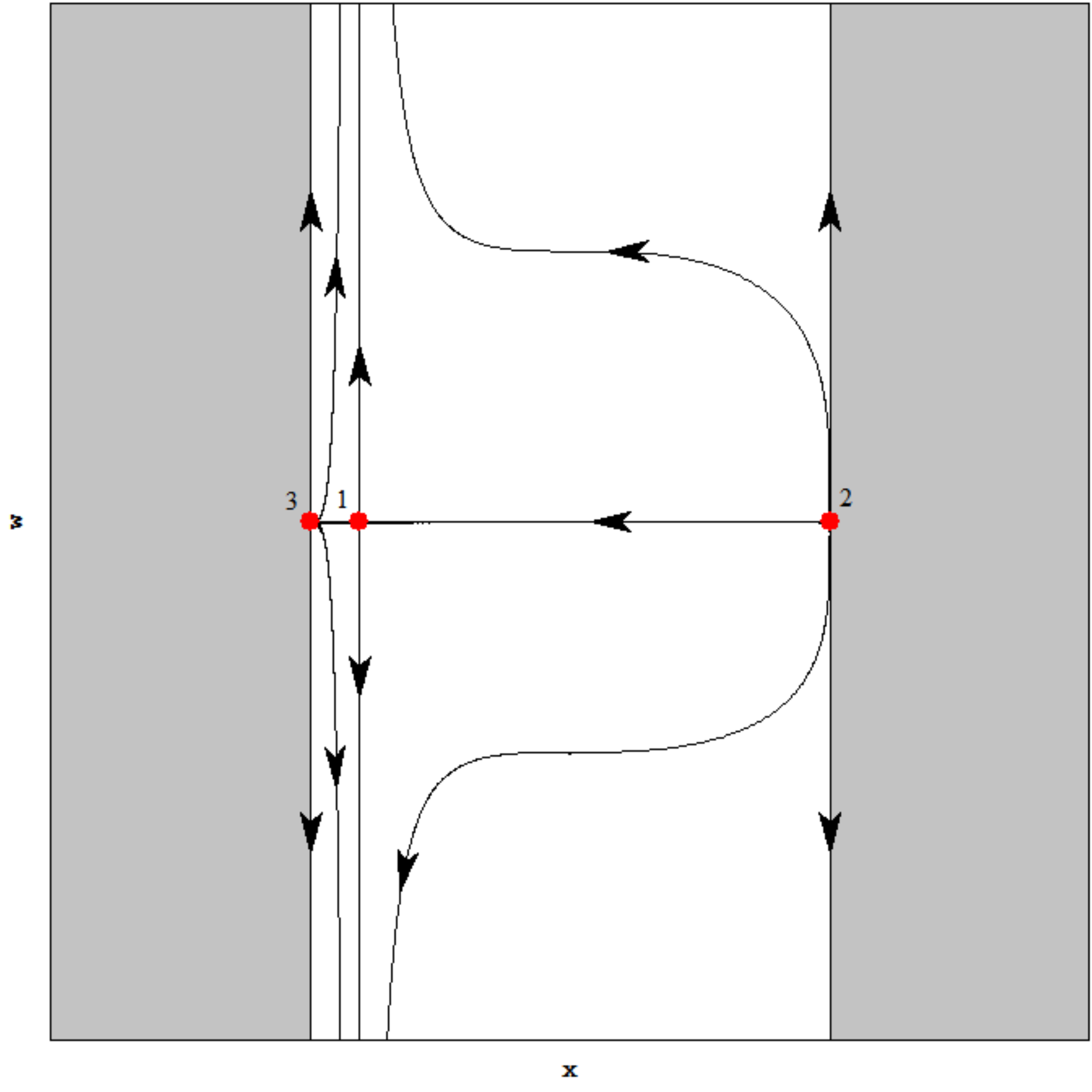}
\caption{A phase portrait for cosmological model (equations \eqref{eq:dyn_exp}) with an exponential potential (for example we put $\lambda=2$) on invariant submanifold ${z=0}$. Critical point (1) is of saddle type. Critical points (2) and (3) are of unstable node types.}
\label{fig:10}
\end{center}
\end{figure}

\begin{figure}
\begin{center}
\includegraphics[scale=0.45]{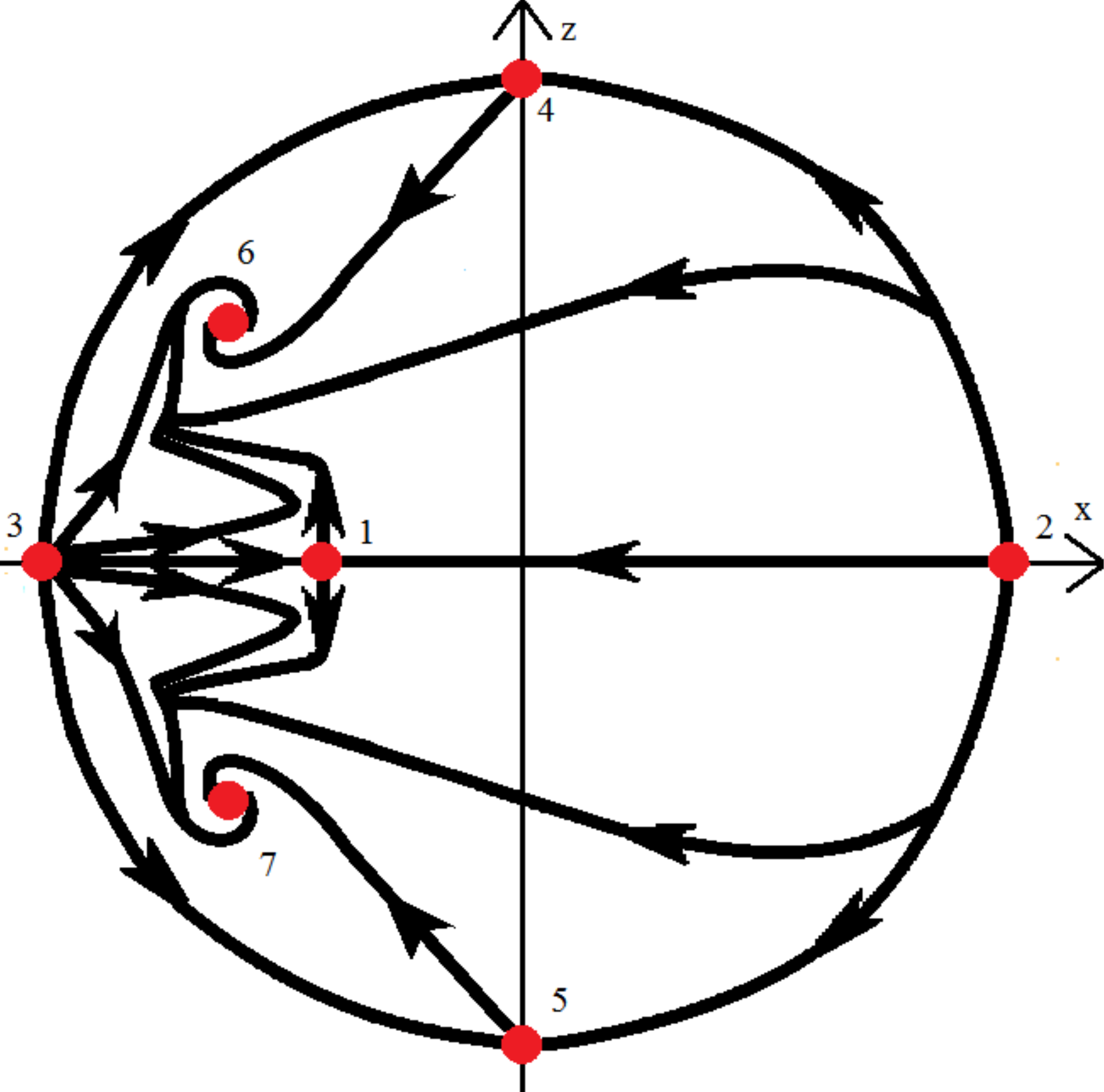}
\caption{A phase portrait for cosmological model (equations \eqref{eq:dyn_exp}) with an exponential potential (for example we put $\lambda=2$) on invariant
submanifold ${w=0}$. Critical points (1), (4) and (5) are of saddle types. Critical points (2) and (3) are of unstable node types. Critical points (6) and (7) are of stable focus types.}
\label{fig:11}
\end{center}
\end{figure}

The system \eqref{eq:dyn_exp} possesses two 2D invariant submanifold $\{z=0\}$, $\{w=0\}$. The phase portraits of the system on these invariant submanifolds are presented in Fig. \ref{fig:10} and Fig.\ref{fig:11}.

Our analysis is general and any special form of potential is not assumed. Scaling solutions are understood in the sense of the same authors \cite{Copeland:1997et}. Energy density of the scalar field $\rho_{\phi}$ is proportional to the energy density of the matter and they both are given in the power law form with respect to the scale factor. In our case the corresponding solutions are represented by lines which start from the critical point $x=x_0,\ z=z_0,\ w=0$ and one of them goes vertically up and the other one down. In this case we obtain from equations \eqref{eq:dyn_exp} and \eqref{eq:hom_sys} the exponential form of a potential function because in this case $W$ is a linear function of $V$.

Finally one can conclude the scaling solution are interesting from the point of view of a solution of the coincidence problem in the quintessence cosmology. However the corresponding evolutional paths have dimension one or zero (a point). The case of the zero dimensional scaling region represents a critical point of the system. Therefore scaling solutions are non-generic because they possess codimension 2.

The role of symmetries in the context of the generation of new cosmological solution for the inflation was investigated in \cite{Liddle:1998xm, Szydlowski:2005nu, Szydlowski:1983}. Liddle and Scherrer classified all forms of the potentials of the scalar field $V(\phi)$ for which the energy density of the scalar field $\rho_\phi=\frac{1}{2}\dot\phi^2+V(\phi)$ scales in a similar way to the barotropic matter as a power law of the scale factor. (They assume that at late time the matter term is dominating over the energy density of the scalar field $\rho_\phi<<\rho_m=\rho_{m,0}a^{-3}$). In the $(\phi, \phi')$ variables and $'=\frac{\ud}{\ud\tau}$, where $\tau=\ln{a}$, a scaling solution appear on the phase plane as a global attractor \cite{Liddle:1998xm}. This condition distinguishes two forms of potentials: a exponential and a power law type $\phi^{-\beta}$ introduced by Ratra and Peebles in the context of the quintessence idea.

There are two ways of a simple use of our knowledge about symmetries:

1) Invariants of a symmetry group are useful candidates for a choice of new state variables and in a consequence for the system inerrability.

2) Invariants enable us to formulate some symmetry theorems which in a consequence can lead to the larger class of special solutions, which form the family of suitable solutions.

Let us illustrate the latter for the case of scaling solutions of the system \eqref{eq:dyn_exp}. If $x(\tau),\ z(\tau)$ and $w(\tau)$ are the solutions of the system which preserve a structure under the action of a homological group: $\tau \rightarrow \bar\tau = h \tau=const$, $x\rightarrow\bar x =ax=const,\ z\rightarrow \bar z=cz=const,\ w\rightarrow\bar w=bw=const$, then invariants are solutions of the equation
\begin{equation}
    \frac{\ud x}{ax}=\frac{\ud z}{cz}=\frac{\ud w}{bw}=\frac{\ud\tau}{h \tau}.
\end{equation}
Here
$\mathcal{I}_1:\ wz^{-\frac{b}{c}}=const$, $\mathcal{I}_2:\ wz^{-\frac{b}{a}}=const$, $\mathcal{I}_3:\ w\tau^{-\frac{b}{a}}=const$
form the base of independent invariants. Therefore if $w(z)$ is a solution of the system then $bw(cz)$ will be also the solution of the same system. Due to homological theorems if some special solutions of the system (for example scaling solutions) are known, then these solutions can be prolonged to the larger class: A family of solutions of the same type.

In many papers scaling solutions are investigated in the so called energetic variables. Such a choice is suggested by structure of energetic constraint. They are in principle projective coordinates if the system is parameterized by the set of state variables
\begin{equation}
    (H^2, \dot\phi^2, V(\phi)).
\end{equation}
Let us consider a flat cosmology with a single scalar field for illustration of this fact. In such a case the dynamics can be written down in the form of the dynamical system
\begin{equation}
\begin{split}
    \frac{\ud}{\ud t}H^2 &=-\dot\phi^2 H\,, \\
    \frac{\ud}{\ud t}(\dot\phi^2) & = -6H\dot\phi^2-2\dot\phi V'(\phi)\,, \\
    \frac{\ud}{\ud t}V &= V'(\phi)\dot\phi\,.
\end{split}
\label{eq:72}
\end{equation}
If to assume that $V'(\phi)=U(V(\phi))$ (this solution is automatically satisfied by the exponential potential) then right-hand sides of the system become homogeneous functions of their arguments if $U(\alpha V)=\alpha U(V)$ (for $V(\phi) \propto e^{\lambda \phi}+const$, the function $U(V)$ is linear, i.e. a homogeneous function of a degree one).

If to define a vector representing the r.h.s. of the system, then $F(x,y,z)=[\mp\sqrt x y,\ \mp6\sqrt x y-2\sqrt y U(z),\ \pm U(z)\sqrt y]^T$ is the vector field in the phase space. Note that for all $\alpha$ we have
\begin{equation}
    F(\alpha x,\ \alpha y,\ \alpha z)=\alpha^{3/2}F(x,y,z).
\end{equation}
Therefore r.h.s. of the dynamical system are homogeneous functions of degree $3/2$. This means that the dimension of the system can be reduced by one by introducing a projective map,
\begin{equation}
    \left(\frac{1}{x},\quad \frac{y}{x},\quad \frac{z}{x}\right).
\end{equation}
Note that three maps cover a circle at infinity and above variables coincides with energetic variables modulo constant value,
\begin{equation}
    \left(\frac{1}{H^2},\quad \frac{\dot\phi^2}{H^2},\quad \frac{V(\phi)}{H^2}\right).
\end{equation}
Finally, the choice of projective coordinates for state variables is very useful because in some situation gives rise to a reduction of the dimension of the system. However it is not possible for all classes of a function of a potential.

Moreover we must remember that the system with any form of $V(\phi)$ is in fact 3D. The reduced system is of course 2D and, in general, the Hubble function $H^2$ can be obtained. The map $(H^2=\infty,\ \frac{1}{6}\frac{\dot\phi^2}{H^2},\ \frac{V(\phi)}{3 H^2})$  covers the circle at infinity $x^2+y^2+z^2=\infty$. Sometimes it may be useful to consider such cases. If we study the behaviour of the system at infinity, i.e. near the singularity then energetic variables will be useful in any case.
If the potential has the exponential form, then 2D reduced system is sufficient for dynamic exploration.

Note that system \eqref{eq:72} is invariant under homological transformation of its state variables. Invariants of these transformation are just chosen as new reduced energetic variables. The set of invariants
\begin{equation}
    \mathcal{I}_1=yx^{-1},\ \mathcal{I}_2=zx^{-1}
\end{equation}
of homological transformations
of the symmetry operator $X=Ax\frac{\partial}{\partial x}+Ay\frac{\partial}{\partial y}+Az\frac{\partial}{\partial z}$ can be chosen as a useful variables for system representation and in consequence useful in the context of system integration.

The Hubble function $H$ appears in the definition of energetic variables due to presence of a normalisation factor. One can ask how this function can be extracted from the trajectories $[x(\tau),\ y(\tau),\ z(\tau)]$ defined in the reduced space. If the cosmological evolution can be fully represented in terms of reduced dynamical system in the energetic variables, then the Hubble function can be derived from trajectories through the simple integration. For illustration of this fact let us consider the case of SFC with the matter. For simplicity we assume flatness of the model. The energetic variables can be defined in a standard way: $x^2=\frac{1}{6}\frac{\dot\phi^2}{H^2}$, $y^2=\frac{V(\phi)}{3H^2}$, $z^2=\frac{\rho_m}{3H^2}$. They satisfy a constraint condition: $x^2+y^2+z^2=1$. After a differentiation of both sides of this condition with respect to the cosmological time $t$ we obtain relation:
\begin{equation}
\label{eq:condif}
    x^2\frac{\ud(\ln x)}{\ud t}+y^2\frac{\ud(\ln y)}{\ud t}+z^2\frac{\ud(ln z)}{\ud t}=0\,,
\end{equation}
where
\begin{equation}
\label{eq:dyn_3}
\begin{split}
    2\ln x=&2\ln\dot\phi-2\ln H+\ln 6^{-1}\,,\\ 
    2\ln y=&\ln(V(\phi))-2\ln H+\ln3^{-1}\,,\\ 
    2\ln z=&\ln\rho_m-2\ln H+\ln 3^{-1}\,.
\end{split}
\end{equation}

Performing a differentiation of Eq. \eqref{eq:dyn_3} with respect to time $t$ and then substituting the result into Eq. \eqref{eq:condif} we obtain the following relation
\begin{equation}
\label{eq:condif2}
    x^2\left(\frac{\ddot \phi}{\dot\phi}-\frac{\dot H}{H}\right)+y^2\left(\frac{1}{2}\frac{V'(\phi)}{V(\phi)}\dot\phi-\frac{\dot H}{H}\right)+z^2\left(\frac{1}{2}\frac{\dot\rho_m}{\rho_m}-\frac{\dot H}{H}\right)=0
\end{equation}
or
\begin{equation}
    \frac{\dot H}{H}(x^2+y^2+z^2)=x^2\frac{\ddot \phi}{\dot \phi}+\frac{1}{2}y^2\frac{V'(\phi)}{V(\phi)}\dot \phi+\frac{1}{2}(-3 H)(1+w_m)z^2.
\end{equation}
Therefore
\begin{equation}
    \frac{\dot H}{H}=x^2\frac{\ddot \phi}{\dot\phi}+\frac{1}{2}y^2\frac{V'(\phi)}{V(\phi)}\dot\phi-\frac{3}{2}H(1+w_m)z^2.
\end{equation}
If to take $\ddot\phi=-3H\dot\phi-V'(\phi)$ from the equation of motion for the scalar field and to substitute this $\ddot\phi$ into the above equation then this equation takes a new form. Defining $\lambda(\phi)=-\frac{V'(\phi)}{V(\phi)}$ Eq. \eqref{eq:condif2} can be rewritten as follows,
\begin{equation}
    \frac{d}{dt}(\ln H)=x^2\left(-3H-\frac{V'(\phi)}{V(\phi)}\frac{V(\phi)}{\dot\phi^2}\dot\phi\right)-\frac{1}{2} y^2\lambda(\phi)\dot\phi-\frac{3}{2}H(1+w_m),
\end{equation}
\begin{equation}
    \frac{1}{H}\frac{d}{dt}(\ln H)=\left(x^2(-3+\lambda(\phi)\frac{2y^2}{x^2}\sqrt{6}x)-\frac{1}{2}y^2\lambda(\phi)\sqrt{6}x-\frac{3}{2}(1+w_m)\right).
\end{equation}
Therefore
\begin{equation}
    \frac{d(\ln H)}{d\tau}=\left(x^2\left(-3+\lambda(\phi)\frac{2y^2}{x^2}\sqrt{6}x\right)-\frac{1}{2}y^2\lambda(\phi)\sqrt{6}x-\frac{3}{2}(1+w_m)\right),
\end{equation}
where $\tau=\ln a$. From the last relation we obtain,
\begin{equation}
    H(\tau)=H_{0}\exp\left(\int^\tau_{0} \left(x^2\left(-3+\lambda(\phi)\frac{2y^2}{x^2}\sqrt{6}x\right)-\frac{1}{2}y^2\lambda(\phi)\sqrt{6}x-\frac{3}{2}(1+w_m)\right)d\tau\right),
\end{equation}
where function $x(\tau)$, $y(\tau)$, $z(\tau)$ are taken from the phase space trajectories and $H_{0}$ is the initial value of the Hubble's function at $\tau=0$.

Finally, information from the reduced phase space about trajectories allows us to obtain a relation for $H(\tau)$ in the model under consideration, i.e. a relation for $H(a)$, which uniquely identifies the FRW cosmological model. The FRW metric for the flat model can be written in the following form: 
\begin{equation}
\ud s^2=a^2 \Big(-\left(\frac{\ud a}{a^{2}H(a)}\right)^{2}+(\ud x^2+\ud y^2+\ud z^2)\Big)\,,
\end{equation} 
and $H^2(a)$ is the only unknown function which uniquely identifies the model.

\section{Conclusions}

Applications of methods of dynamical system and investigations of the Lie symmetry in the context of cosmological dynamical systems were the main subject of our investigations in this paper. We have used these methods to cosmological models with scalar field and the matter. We have shown how all scalar field cosmological models can be equipped with the structure of the Banach space. In our considerations different cosmological models are represented by different potentials of the scalar field and we can measure the distance between them. We have found the scalar field cosmological model which reproduces the Standard Cosmological Model ($\Lambda$CDM model). The $\Lambda$CDM models are structurally stable (and therefore they are generic in the ensemble), and therefore, the same property is inherited by the scalar field cosmological model.

We have also studied the phase space of scalar field cosmological models with a non-minimal coupling to gravity. It was demonstrated how some structurally unstable models with non-minimally coupling bifurcate to structurally stable ones. Such a mechanism is sensitive to the choice of the potential function.

Recent Planck satellite observations \cite{Ade:2013rta} has eliminated wide range of complex inflationary scenarios and favour simple models with one scalar field with potential. Astronomical observations favour inflationary models with a potential function with wide plateau. In such a case the potential of the inflaton behaves as $V(\phi)\approx V_{0} = \text{const}$ and the corresponding system is structurally stable.

The Peixoto theorem characterises generic models in the ensemble in the case of a 2-dimensional dynamical system. They form open and dense subsets in the ensemble. Our point of view is that such models are extremely interesting in the scalar field cosmology in the context of solving of the indetermination problem of the scalar field potential. We have also characterised a class of potentials of the scalar field which give rise to structurally stable cosmology.

The scaling solution plays an important role in the modern cosmology. They are interesting when explaining why densities of both dark energy and the matter are comparable (of the same order) in the current Universe. A traditional understanding of a scaling solution comes from Copeland et al. \cite{Copeland:1997et}. Our understanding is more general because we define homological symmetries of the basic equation instead of a scaling solution and we are looking for such domains in the phase space where such symmetries are obvious. We show that interesting from the point of view solution of coincidence problem, scaling solutions are not global attractor and forms only the set of zero measure in the phase space.

\section*{Acknowledgments}
We are very grateful K. Urbanowski and A. Krawiec for proofreading the manuscript and for the interesting comments.
The research of OH was supported by the
National Science Centre through the post-doctoral internship award (Decision
No.~DEC-2012/04/S/ST9/00020).

\bibliographystyle{JHEP}
\bibliography{sfc_geometry}

\end{document}